\documentstyle[eqsecnum,aps,epsf]{revtex}

\begin{document}
\draft

\newcommand {\beq}{\begin{eqnarray}}
\newcommand {\eeq}{\end{eqnarray}}
\newcommand {\be}{\begin{equation}}
\newcommand {\ee}{\end{equation}}
\newcommand {\e}{\epsilon}
\newcommand{\ds}[1]{#1 \hspace{-5.2pt}/} 
\newcommand{\Ds}[1]{#1 \hspace{-1ex}/} 
\def\bfm#1{\mbox{\boldmath $#1$}}
\newcommand{\BQ}{\begin{equation}}
\newcommand{\EQ}{\end{equation}}
\newcommand{\BQA}{\begin{eqnarray}}
\newcommand{\EQA}{\end{eqnarray}}
\newcommand{\half}{\frac{1}{2}}
\newcommand{\NN}{\nonumber \\}
\newcommand{\E}{{\rm e}}
\newcommand{\Gmu}{\gamma^{\mu}}
\newcommand{\Gnu}{\gamma^{\nu}}
\newcommand{\gmu}{\gamma_{\mu}}
\newcommand{\gnu}{\gamma_{\nu}}
\newcommand{\gfive}{\gamma_5}
\newcommand{\del}{\partial}
\newcommand{\k}{\mbox{\boldmath $k$}}
\newcommand{\itl}{\mbox{\boldmath $l$}}
\newcommand{\itP}{\mbox{\boldmath $P$}}
\newcommand{\q}{\mbox{\boldmath $q$}}
\newcommand{\p}{\mbox{\boldmath $p$}}
\newcommand{\x}{\mbox{\boldmath $x$}}
\newcommand{\y}{\mbox{\boldmath $y$}}
\newcommand{\Z}{{\bf Z}}
\newcommand{\R}{{\rm R}}
\newcommand{\M}{{\cal M}}
\newcommand{\dagg}{\mbox{\scriptsize{\dag}}}
\newcommand{\dagpsi}{\psi^{\mbox{\scriptsize{\dag}}}}
\newcommand{\ket}[1]{\left.\left\vert #1 \right. \right\rangle}
\newcommand{\bra}[1]{\left\langle\left. #1 \right\vert\right.}
\newcommand{\ketrm}[1]{\vert {\rm #1} \rangle}  
\newcommand{\brarm}[1]{\langle {\rm #1} \vert}  

\topmargin=0cm

\title{Structural Change of  Cooper Pairs and 
 Momentum-dependent Gap \\ in Color Superconductivity}

\author{Hiroaki Abuki\footnote{\tt abuki@nt.phys.s.u-tokyo.ac.jp} and
Tetsuo Hatsuda\footnote{\tt hatsuda@phys.s.u-tokyo.ac.jp}}
\address{Department of Physics, University of Tokyo, Tokyo 113-0033,
Japan}

\author{Kazunori Itakura\footnote{\tt itakura@bnl.gov}}
\address{RIKEN BNL Research Center, Brookhaven National Laboratory,
              Upton, NY 11973, USA}

\date{\today}
\maketitle

\begin{abstract}

 The two-flavor color superconductivity is studied 
   over a wide range of baryon density with a single model.  
   We pay a special attention
   to  the {\it spatial}-momentum dependence of the gap 
   and to the {\it  spatial}-structure of Cooper pairs.
At extremely high baryon density ($\sim {\cal O}(10^{10} \rho_0)$
   with $\rho_0$ being the normal nuclear matter density),
   our model becomes equivalent to the usual perturbative QCD treatment
   and  the gap is shown to  have  a sharp peak near the Fermi surface  
   due to the weak-coupling nature of QCD.
 On the other hand,  the gap is a smooth function of the
   momentum  at lower densities ($\sim {\cal O}(10 \rho_0)$)
  due to  strong color magnetic and electric interactions.
 To study the structural change of Cooper pairs
   from high density to lower density, quark correlation
   in the color superconductor is studied both in the 
   momentum space and in the coordinate space. 
 The size of the Cooper pair is shown to become comparable
   to the averaged inter-quark distance at low densities.
 Also, effects of the momentum-dependent running coupling
    and the antiquark pairing, which are both small
    at high density, are shown to be non-negligible at low densities. 
  These features are highly contrasted to the standard BCS
    superconductivity in metals.
\end{abstract}


\section{Introduction}

Because of the asymptotic freedom and the Debye screening in QCD,
 deconfined quark matter is expected to be realized 
 for baryon densities much larger than the normal nuclear matter
 density \cite{Collins_Perry}. Furthermore, any attractive quark-quark 
 interaction in the cold quark matter  causes an instability of the Fermi 
 surface by the formation of Cooper pairs and leads  to 
 the color superconducting phase
 \cite{BL_84,Iwasaki,ARW_98,RSSV_98,Review}.

 Current understanding on the color superconductivity has been
 based on two different theoretical approaches. One is an analysis
 by the one-loop Schwinger-Dyson  equation with perturbative one gluon 
 exchange, which is valid at asymptotically 
 high densities \cite{Schaefer,Hong,Pisarski-Rischke}.  
 The dominant contribution to the superconducting gap
 at very high density comes from the collinear scattering through the 
 long range magnetic gluon exchange \cite{Son_98}.
 In such a weak-coupling regime,  formation of 
 Cooper pairs takes place only in a small region near the Fermi surface. 
 The other approach to the color superconductivity 
  is the mean-field approximation with QCD inspired 4-Fermi 
  model which is introduced to study lower density 
  regions\cite{ARW_98,RSSV_98,Berges-Rajagopal,Diakonov}.
 In this approach, magnitude of the gap becomes as large as 100 MeV and 
  is almost constant in the vicinity of Fermi surface as a 
       function of momentum.

 The main purpose of this paper is to calculate the superconducting
 gap over a wide range of baryon density with a single model and 
 to make a bridge between high and low density regimes.
 In particular, we make an extensive analysis on the structural 
 change in spatial-momentum dependence of Cooper pair from high to low 
 densities. At relatively low baryon densities, the gluonic 
 attraction becomes effective for all quarks inside the Fermi sea and 
 sizable Cooper pairing takes place for wide range of momentum 
  away from the Fermi surface.
 This is in contrast to the phonon interaction
 with an intrinsic Debye cutoff in the BCS-type superconductivity\cite{BCS}. 
 The {\em spatial}-momentum dependence of the gap, diffuseness of 
 the Fermi surface, quark-quark correlations in the superconductor and the 
 spatial size of Cooper pairs are the characteristic quantities which 
 reflect the departure from the weak-coupling picture. 
 So far, only a few investigations have been made along this line
 \cite{Iwasaki,Horie,Matsuzaki}.

Throughout the paper, we limit our discussions to 2-flavor color
 superconductivity partly because our primary interests are in low 
 densities and partly because the analysis is simpler than in the 
 three-flavor case \cite{ARW_98b}. 
 To investigate how the weak-coupling picture is modified at low 
 densities in a qualitative manner,  we adopt a model 
 called the improved ladder approach in which the one-loop Schwinger-Dyson 
 equation with infrared safe running coupling is used. This model is known 
 to reproduce the physics of the QCD vacuum and the $q\bar{q}$ meson 
 properties reasonably well \cite{Higashijima,Aoki,Haymaker}.
 For recent applications of similar kind of
  models to QCD at finite temperature   
 and density, see the review \cite{Roberts}.
 As we will show later, the model can naturally
 reproduce the correct asymptotic behavior of the superconducting gap,
 in the high density limit. Therefore, it is a suitable model
 for our purpose to study the color superconductivity 
 for a wide range of density.
 To make our analysis as simple as possible,
  we use a gluon propagator with a minimal static and
  dynamical corrections (Debye screening and Landau damping).
 The similar form of the gluon propagator is also adopted in Ref.~\cite{Iida}.

This paper is organized as follows. 
 In Sect.~II, we define our model and 
  derive relevant gap equations. We first consider the weak-coupling region 
  at high density and discuss the general properties of the gap as a function 
  of the spatial-momentum.
 In Sect.~III, we solve the momentum-dependent gap equation
  numerically for a wide range of density. A large structural change of 
  the momentum-dependent gap will be shown. The quark and antiquark 
  occupation numbers, the correlation of quarks in the color superconductor,
  and the coherence length are also investigated.
 Summary and concluding remarks are given in the last section.
 In Appendix A, the gauge dependence of the gap equation is discussed.
 In Appendix B, we derive the asymptotic behavior of the spatial quark-quark
  correlation in the weak coupling.

\section{Gap equation with spatial momentum dependence}

In this section, we define our model and derive relevant gap equations in
  two flavor QCD with the one-gluon exchange interaction.
We start with the standard Schwinger-Dyson equation for the
  quark self-energy  with four momentum dependence \cite{Schaefer}.
Then it is reduced to a  gap equation with spatial-momentum dependence.
General properties of the solution of our gap equation
  in the  weak-coupling limit   are also examined.
The gap  with spatial-momentum dependence enables us to study the
   diffuseness of the Fermi surface and the 
   the  structure of Cooper pairs in the color superconductor.

Throughout this paper, we use the following notation for the
 four and three momenta : $k^{\rho} = (k^0, \k ),\,k = | \k |,%
 \,\hat{\k} = \k /k.$
Also, we limit ourselves to the system at zero temperature and 
 work in the Minkowski space with the quark and gluon propagators 
    obeying the Feynman boundary condition.

\subsection{The model}

Let us start with the definition of quark self-energy and the
  superconducting gap.
Using the standard Nambu-Gor'kov formalism with a two component Dirac
spinor
  $\Psi=(\psi,\bar{\psi}^t)$, the quark self-energy $\Sigma(k_{\rho})$
 with the Minkowski 4-momentum $k_{\rho}$  is written as
\begin{eqnarray}
  \Sigma(k_{\rho})=S_0^{-1}(k_{\rho})-S^{-1}(k_{\rho})&=&\left(%
\matrix{%
0 & \bar{\Delta}(k_{\rho})  \cr
\Delta(k_{\rho})  & 0 \cr
}
\right),
\end{eqnarray}
where the superconducting gap $\Delta$ and
  $\bar{\Delta}\equiv\gamma_0\Delta^\dagger\gamma_0$
  enter through the off-diagonal components of $\Sigma$.
The diagonal components of $\Sigma$ is neglected throughout this paper
  \cite{Schaefer}.
$S(k_{\rho})$ and $S_0(k_{\rho})$ are the full and free quark propagators,
  respectively.
The free propagator is taken to be a form $S_0^{-1}(k_{\rho}) =
   {\rm diag.}(k\!\!\!/ +\mu\!\!\!/ , (k\!\!\!/ -\mu\!\!\!/)^T )$.
We ignore the mass of $u$ and $d$ quarks.

In the ladder approximation of the one gluon exchange,
 the Schwinger-Dyson equation for $\Sigma$ is written as
\beq
  \Sigma(k_{\rho})&=&-i \int\frac{d^4q}{(2\pi)^4} \ g^2 \ 
 \Gamma_a^\mu S(q_{\rho})%
  \Gamma_b^\nu D_{\mu\nu}^{ab}(q_{\rho}-k_{\rho}),\label{DSeq}
\eeq
where $D^{ab}_{\mu\nu}$ is the gluon propagator in medium and
 $\Gamma^a_\mu$ is the quark-gluon vertex.

To study the flavor anti-symmetric, color anti-symmetric and
  $J=0^+$ channel 
(which is the most attractive channel within the one gluon exchange model),
 we assume the following structure of the gap function
 \cite{Schaefer,Pisarski-Rischke,Pisarski-Rischke2},
\begin{equation}
\Delta(k_{\rho})=(\lambda_2 \tau_2 C\gamma_5)
 \left( \Delta_+(k_{\rho})\Lambda_+(\hat{\k} ) +
    \Delta_-(k_{\rho})\Lambda_-(\hat{\k} ) \right) .
\end{equation}
Here $\tau_2$ is the Pauli matrix acting on the flavor space, 
$\lambda_2$ is a color anti-symmetric Gell-Mann matrix,
and  $C$ is the charge conjugation. $\Lambda_\pm(\hat{\k})\equiv%
(1\pm\hat{\k}\cdot\bfm{\alpha})/2$ is the projector on 
 positive ($+$) and negative $(-)$ energy quarks.
$\Delta_\pm$ will be identified as the gap in quark and 
  antiquark channel, respectively.

For the spinor and color structures of the 
 vertex in Eq.~(\ref{DSeq}), we use the bare vertex
$\Gamma^a_\mu=\mbox{diag.}(\gamma_\mu \lambda^a/2, %
 -(\gamma_\mu \lambda^a/2)^t)$ \cite{Schaefer}.
For $g^2$ in Eq.~(\ref{DSeq}), we use a
 momentum dependent coupling $g^2(q,k)$ 
 which does not diverge at low momentum scale.
  In the ``improved ladder approximation'' \cite{Higashijima},
 $g^2(q,k)$ is taken to be
\beq
  g^2(q,k)=\frac{8\pi^2}{\beta_0} \frac{1}{\ln \left(%
  (p^2_{\rm max} +p_c^2)/{\Lambda^2} \right)},
  \quad p_{\rm max}={\rm max}(q,k),
  \label{HM}
\eeq
where $\beta_0=(11N_c-2N_f)/3$, $p_c^2$ plays a role of  a  phenomenological
 infrared regulator, and $\Lambda$ dictates the logarithmic decreases of 
 the vertex at high momentum. The quark propagator in the improved ladder 
 approximation in the vacuum is known to have a high momentum behavior 
 consistent with that expected from the  renormalization group and the 
 operator product expansion. For the numerical values of 
 $\Lambda$ and $p_c^2$, we adopt 400 MeV and 1.5 $\Lambda^2$ respectively.
 They are determined to reproduce the low energy meson properties for $N_f=2$
 by solving the Schwinger-Dyson equation for the quark propagator and 
 the Bethe-Salpter equation for the $q\bar{q}$ bound state in the 
 vacuum~\cite{Aoki}.

We use the following gluon propagator in medium  in Eq.~(\ref{DSeq}) 
in the Landau gauge,
\beq
  D_{\mu\nu}(k_{\rho})=
-\frac{P^{\rm T}_{\mu\nu}}{\bfm{k}^2+iM^2 |k_0|/|\bfm{k}|}%
   -\frac{P^{\rm L}_{\mu\nu}}{\bfm{k}^2+m_{\rm D}^2}, \label{propagator}
\eeq
where $P^{\rm T,L}_{\mu\nu}$ are the transverse and longitudinal
projectors satisfying the relations,
 $P^{\rm T}_{ij}=\delta_{ij}-\hat k_i \hat k_j $, 
$ P^{\rm T}_{00}=P^{\rm T}_{0i}=0$, and $P^{\rm L}_{\mu\nu}%
=-g_{\mu\nu}+{k_\mu k_\nu}/{k_{\rho}^2}-P^{\rm T}_{\mu\nu}$. 
 The longitudinal part of the propagator has static 
 screening by the Debye mass
 $m_{\rm D}^2 = (N_f/2\pi^2) g^2 \mu^2 $, while the
 transverse part has dynamical screening by the 
 Landau damping $M^2=(\pi/4)m_{\rm D}^2$ \cite{LeBellac}.
 The above form of the propagator is a quasi-static
 approximation of the full gluon propagator in the sense
 that only the leading frequency dependence is considered.
  The same form is also adopted in Ref.~\cite{Iida}.

Here we make a brief comment on the gauge-parameter $\xi$ dependence 
 of the gap equation. Although physical quantities should be gauge invariant, 
 special truncation scheme of the diagrams such as the ladder approximation 
 introduces gauge-parameter dependence. It has been claimed that
 (i) $\Delta_+$ is gauge invariant at extremely high density \cite{Schaefer},
and  (ii) the gauge dependent contribution only begins to
 decrease at extraordinarily high densities $\mu > 10^8$MeV
 and seems to converge to the result of the Landau gauge ($\xi=0$)%
 \cite{gauge_dependence}. We will show in Appendix A that, 
if one appropriately treats the momentum dependence,
 the $\xi$-dependence survives in the 
 gap equation for $\Delta_+$ but gives a sub-leading contribution.  
 Under this remark, we will consider only the Landau gauge ($\xi =0$) 
 throughout this paper.

The gap equation under the approximations shown above
 is obtained from the 2-1 element of the Schwinger-Dyson 
 (matrix) equation (\ref{DSeq}).
Then the final form of the gap equation for $\Delta_\pm(k_{\rho})
 = \Delta_\pm(k_0,  \k )$ reads
\beq
  \Delta_\pm(k_{\rho})&=&\frac{N_c+1}{2N_c} i
  \int\frac{d^4q}{(2\pi)^4}g^2(q,k)%
  \left\{\frac{1}{2}{\rm tr}\left(\Lambda_\pm(\hat{\bfm{k}})\gamma^\mu%
  \Lambda_-(\hat{\bfm{q}})\gamma^\nu\right)%
  \frac{\Delta_+(q_{\rho})}
         {q_0^2-E_+^2(q)-|\Delta_+(q_{\rho})|^2}\right.\nonumber \\
    & &\ +\left.\frac{1}{2}{\rm tr}\left( \Lambda_\pm(\hat{\bfm{k}})%
  \gamma^\mu \Lambda_+(\hat{\bfm{q}})\gamma^\nu\right)
        \frac{\Delta_-(q_{\rho})}%
  {q_0^2-E_-^2(q) -|\Delta_-(q_{\rho})|^2}\right\}
        D_{\mu\nu}(q_{\rho}-k_{\rho}), \label{gap_4}
\eeq
where $E_\pm(q) (= q \mp \mu)$ is the reduced energy for
 free quarks (for $+$ sign) and for antiquarks (for $-$ sign).

\subsection{Analytic properties of the gap in the weak coupling}

\subsubsection{Gap equation in the weak-coupling limit}

 Let us first study the analytic properties of the gap equation with 
    spatial-momentum dependence. 
 To obtain analytically tractable equations, we consider
   here the weak-coupling (high density) limit.
 This ``weak-coupling approximation'' implies the approximation where
   we take only the color-magnetic interaction and the real part of
   $\Delta_+$. Momentum dependence of $g^2(q,k)$ is  
 neglected.  Contributions from the antiquarks proportional to
 $\Delta_-$ and the momentum-dependent factor originating from the  
  projection operator in Eq.~(\ref{gap_4}) are also neglected.
We also ignore the imaginary part of the propagator because it is 
   factor $g^2$ smaller than the real part. 
  In Eq.~(\ref{gap_4}), the frequency integral is performed by picking 
  up quasi-particle poles in the quark propagator.
 After integrating over the angular variable, we obtain the following 
  gap equation with the spatial-momentum dependence
\beq
  \Delta_+(k)= \frac{g^2 }{144\pi^2}\int_0^\infty  dq%
\frac{\Delta_+(q)}%
{\sqrt{E_+^2(q)+\Delta_+^2(q)}}\left(\frac{q}{k}%
\ln\left[\frac{(q+k)^6+M^4(\e^+_k-\e^+_q)^2}{(q-k)^6%
+M^4(\e^+_q-\e^+_k)^2}\right]+(\e^+_q\rightarrow %
-\e^+_q)\right), \label{gap_simple1}
\eeq
where $E_+(q) \equiv q - \mu$ and we have defined 
 $\Delta_+(q) \equiv  \Delta_+ (\e_q^+,q)$
 for notational simplicity. 
 $\e_q^+$ is the quasi-particle energy
 as a solution of $(\e_q^+)^2 = 
        E_+^2(q) + \Delta_+^2(\e_q^+,q)$.
This is the final form of an approximate gap equation in the 
  weak-coupling limit.
  $\Delta_+(k)$ corresponds to {\it the gap of the
    on-shell quasi-particles} in the color superconductor
  since $k_0$ is taken to be $\e_k^+$.

\subsubsection{Gap near the Fermi surface in weak coupling limit}

Let us consider the behavior of the gap near the Fermi surface.
As we claimed in Introduction, the gap equation (\ref{gap_simple1}) 
 reproduces the correct high density solution near the Fermi surface,
 which was previously 
 derived in Refs.~\cite{Schaefer,Son_98} for frequency dependent gap,
 and in Ref.~\cite{Pisarski-Rischke} for momentum dependent gap.
 For this purpose, we pick up the most singular
 part of the integrand by taking $q=k=\mu$ in the bracket except 
  for the term  $(\e_q^+ -\e_k^+)^2$ in the denominator.
 Then Eq.~(\ref{gap_simple1}) reduces to
\beq
   \Delta_+(k\sim \mu )= \frac{g^2}{72\pi^2}%
    \int_{0}^{\infty} dq%
\frac{\Delta_+(q)}{\sqrt{E_+^2(q)+|\Delta_+(q)|^2}}%
  \ln \frac{(b  \mu)^2}{\left|\e^{+2}_q-\e^{+2}_k\right|},\label{noDFS}
\eeq
where   $b = 2 (2 \mu/ M)^2 = (2/N_f)(32 \pi/g^2) $.
 This equation makes sense only in the vicinity of the Fermi
 surface $|k-\mu|\ll\delta$ and coincides to the one derived 
 in Ref.~\cite{Pisarski-Rischke}.

In the weak-coupling regime ($g/(3\sqrt{2}\pi ) \ll 1$),
  the approximate solution of this equation is known to be
     \cite{Pisarski-Rischke}
\beq
  \Delta_+(k \sim \mu )
 \cong \Delta_+ (\mu)
        \cos\left(%
        \frac{g}{3\sqrt{2}\pi}\ln \frac{|k-\mu|+\e_k^+}%
        {\Delta_+(\mu)}\right),
\ \ \    \Delta_+ (\mu) \cong  2 b  \ \mu\ %
        e^{-(3\pi^2/\sqrt{2})/g},
\label{sol-LD}
\eeq 
where $\e_k^+\cong (|k-\mu|^2+\Delta^2_+(\mu))^{1/2}$.
This solution is a symmetric function of $(k-\mu)$ and decreases as
$|k-\mu|$ increases. 
Inclusion of the  color-electric interaction only modifies 
  the prefactor as
  $b \to  (2/ N_f  )^{5/2} 256 \pi^4/g^5$,  which 
  enhances the gap in the weak coupling.
The characteristic form of the gap $e^{-c/g}$ in
  Eq.~(\ref{sol-LD}), which is not the BCS form $e^{-c/g^2}$, 
  was first derived in Ref.~\cite{Son_98}.

\subsubsection{Gap far from the Fermi surface in weak coupling limit}

Next, we examine the behavior of the gap far from the Fermi surface.
 When $k\gg \mu$, the gap equation (\ref{gap_simple1}) can be cast into the
  following differential equation
\beq
k\frac{d^2}{dk^2}\Delta_+(k)+3\frac{d}{dk}\Delta_+(k)
  +\frac{g^2}{3\pi^2}\frac{1}{k}\Delta_+(k) =0,
\eeq
where we have assumed that $\Delta_+(k)\rightarrow 0$ as
  $k\rightarrow \infty$.
The asymptotic solution of this equation for $g^2 < 3\pi^2$
 is
\beq
\Delta_+(k \gg \mu )\propto k^{-\lambda} \label{power},
\ \ \ \lambda=1+\sqrt{1-g^2/3\pi^2}.
\eeq
Therefore,  in the  weak coupling $g \ll 1$, the gap decreases slowly 
 as $1/k^2$ for $k \gg \mu$    and does not vanish for arbitrary large $k$.
This long tail is a direct consequence of the gluonic interaction which 
 allows scattering of quarks with an arbitrary energy-momentum transfer. 
 Otherwise, the gap cannot take the nonzero value far away from the Fermi 
 surface.

This is in sharp contrast to the gap in the standard weak-coupling BCS 
 superconductivity in metals, %
 \footnote{This is a solution of the simplest gap equation discussed in the 
 original BCS paper \cite{BCS}
$$
\Delta^{\rm BCS}(k)=\frac12 \sum_{k'}V_{k,k'}
\frac{\Delta^{\rm BCS}(k')}{\sqrt{(\e_k-\mu)^2+\Delta^{\rm BCS}(k')^2}}
$$
where $V_{k,k'}\neq 0 $ for
$\e_{\rm F}-\omega_{\rm D} <\e_{k,k'}<\e_{\rm F}+\omega_{\rm D}$ and
$V_{k,k'}=0$, otherwise.
}
where
\be
\Delta^{\rm BCS}(p)=
\left\{
\matrix{
\Delta & \mbox{ $ |\e_p - \e_{\rm F} | < \omega_{\rm D} $}\cr
0      & \mbox{\hspace{-0.5cm} otherwise,}\cr
}
\right.\label{BCS}
\ee
with $\epsilon_p$, $\epsilon_{\rm F}$ 
 and $\omega_{\rm D}$ 
being the electron energy, the electron
 Fermi energy and the ultraviolet 
 Debye cutoff of the phonon interaction originating from the 
 lattice structure, respectively. As we mentioned above, the long tail 
 in Eq.~(\ref{power}) is intimately related to the absence of the 
 {\em intrinsic} ultraviolet-cutoff such as $\omega_{\rm D}$ in QCD.


\subsection{Full gap equation}

 Now let us derive the full gap equation with momentum dependence 
  from the original gap equation  (\ref{gap_4}) with 4 momentum dependence.

In Eq.~(\ref{gap_4}), we adopt the static gluon propagator
 with the Debye screening (for the electric part) and
 the Landau damping (for the magnetic part).
 The imaginary part of the gap function will be neglected as
 we stated before. 
Using the approximation
  $P^{\rm L}_{\mu\nu}\approx -\delta_{\mu 0}\delta_{\nu 0}$ and performing 
  the $q_0$ and angular integration, the gap equation becomes
\beq
\Delta_\pm(k) =
         \int_0^\infty dq\  V_\pm(q,k;\e_q^+,\e_k^{\pm})
           \frac{\Delta_+(q)}{\sqrt{E_+(q)^2+|\Delta_+(q) |^2 }}
        +\int_0^\infty dq\  V_\mp(q,k;\e_q^-,\e_k^{\pm})
           \frac{\Delta_-(q)}{\sqrt{E_-(q)^2+|\Delta_-(q) |^2 }},
      \label{full-gap}
\eeq
where we used a simplified notation: 
$\Delta_{\pm}(\e_k^{\pm},k) \rightarrow \Delta_{\pm}(k)$,
and $V_\pm$ is defined by
\beq
V_\pm(q,k;q_0,k_0)&=&\frac{g^2(q,k)}{48\pi^2}\frac{q}{k}
             \biggl[\frac{1}{3}\ln \left(\frac{(k+q)^6%
             +M^4(q_0-k_0)^2}{(k-q)^6+M^4(q_0-k_0)^2}\right)%
             +(q_0\rightarrow-q_0)+f_{\rm MG}\biggl]\NN
    &+&\frac{g^2(q,k)}{48\pi^2}\frac{q}{k}%
            \biggl[\frac{(k\pm q)^2 + m_{\rm D}^2}{2qk}
            \ln \left(\frac{(k+q)^2+m_{\rm D}^2}{(k-q)^2+m_{\rm D}^2}
                 \right)+f_{\rm EL}\biggl].\label{full-gap-int}
\eeq
 Here $f_{\rm EL}$ is just a constant, and $f_{\rm MG}$ is a 
 rather complicated function of $q$ and $k$.
 Since no singularity appears in $f_{\rm EL,MG}$ 
 when $m_{\rm D} =0$, $M=0$ and  $q=k=\mu$ are taken, 
 they are considered to be sub-leading contribution and 
 will be neglected in the following analysis.  
In the definition of $V_\pm $, the first (second) term corresponds 
to the magnetic (electric) gluon  contribution.
We call the first integral in Eq.~(\ref{full-gap}) as  ``quark-pole
contribution'' and the second integral as ``antiquark-pole contribution'',
because $E_+ = q - \mu$ ($E_- = q + \mu$) is involved in the former (latter).
Near the Fermi surface $q\sim \mu$, the first integral
  in Eq.~(\ref{full-gap})  gives the dominant
  contribution due to the small denominator $(E_+^2+|\Delta_+|^2)^{1/2}=
  \{(q-\mu)^2 +|\Delta_+|^2\}^{1/2}\sim 0$.

At extremely high density, the Cooper pairing is expected to take
  place only near the Fermi surface as we have discussed.
 In this case,  we can safely neglect the antiquark-pole contribution for
 calculating $\Delta_+$.
 Furthermore, one may replace the momentum dependent vertex by the
  coupling constant on the Fermi surface
 $ g^2(q,k) \rightarrow g^2(q=\mu,k=\mu).$  
If $\mu$ is large enough, $g^2(q=\mu,k=\mu)$ may be
 identified with the standard running coupling
 $g^2(\mu)=(8\pi^2 /\beta_0 )/{\ln ( \mu^2/ \Lambda_{\rm QCD}^2 )}$.
 Thus, keeping only the magnetic interaction, one recovers the
  gap equation (\ref{gap_simple1}) for $\Delta_+$
   with $g^2 \rightarrow g^2(\mu) $
    in the weak-coupling limit.

At low densities,
  sizable diffusion of the Fermi surface occurs and the
  weak-coupling  approximations leading to
  Eq.~(\ref{gap_simple1}) are not justified.
 Therefore, we need to solve
  the coupled gap equations  Eq.~(\ref{full-gap}) numerically.
 In particular, 
 the replacement $ g^2(q,k) \rightarrow g^2(q=\mu,k=\mu)$ is not 
 justified  when $q$ and $k$ are not close to $\mu$.
 The contribution from the antiquark-pole is also
 not entirely negligible.

\subsection{Occupation number, correlation function and coherence length}

To clarify the structural change of the color superconductor
 from high to low densities,
 it is useful to examine the following  physical quantities
 related to the gap function.

\vspace{0.2cm}

\noindent
(i)   The quark and antiquark occupation number
   in  each momentum state: This is 
   a fundamental quantity characterizing the 
  diffuseness of the Fermi surface. 
 It is related to the diagonal (1-1) element of the 
 quark propagator in the   Nambu-Gor'kov formalism
\beq
&&\hspace{-0.5cm}
 \langle\psi^{\dagger b}_{j}(t,\bfm{y})\psi^a_i(t,\bfm{x})\rangle_{\rm
super}
=\lim_{x^0\to y^0-\epsilon}
 \left(-i [S_{11}]_{ij}^{ab}(x-y)\gamma^0\right) \nonumber\\
  &&=\int\frac{d^3\bfm{q}}{(2\pi)^3}%
     e^{i{\scriptsize \bfm{q}(\bfm{x}-\bfm{y})}}\Biggl[%
     \biggl\{\Lambda_+(\hat{\q})
       \theta(\mu-|\bfm{q}|)+ \Lambda_-(\hat{\q})\biggl\}
            (P_3^c)_{ab}({\bf 1}_{\rm F})_{ij}\nonumber\\
  &&\quad +\biggl\{%
      \frac{1}{2}\biggl(1%
   -\frac{E_+(q)}{\sqrt{E_+(q)^2+|\Delta_+(q)|^2}}\biggl)\Lambda_+%
  + \frac{1}{2}\biggl(1%
     +\frac{E_-(q)}{\sqrt{E_-(q)^2+|\Delta_-(q)|^2}}\biggl)\Lambda_-%
     \biggl\}(1-P_3^c)_{ab}({\bf 1}_{\rm F})_{ij}\Biggl],
\eeq
where $P_3^c$ is a projection matrix to the third axis in the
 color space.
 From this expression, one can extract the  quark and the antiquark
 occupation numbers as
\be
n_{\pm}^{1,2}(q)
=\frac{1}{2}\left(1-\frac{E_{\pm}(q)}{\sqrt{E_{\pm}(q)^2
 +|\Delta_{\pm}(q)|^2}}\right),\quad
n_+^3(q)=\theta(\mu- q), \quad
n_-^3(q)=0,
 \label{occupation}
\ee
where the superscripts (1,2 and 3) stand for color indices.
Since the third axis in the color space
  is chosen to break the color symmetry,  quarks
 with  the third color do not contribute to form Cooper pairs.
 When the gap is zero $\Delta_\pm=0$, the system
 reduces to the ordinary  quark matter
 with a sharp Fermi sphere;
 $n_+^{1,2}(q)=\theta(\mu-q)$ and $n_-^{1,2}(q)=0$.

\vspace{0.2cm}

\noindent
(ii)   The
 $q$-$q$ and $\bar{q}$-$\bar{q}$ correlation functions in the 
   momentum space $\hat{\varphi}_{\pm}(q)$
  and in the coordinate
 space ${\varphi}_{\pm}(r)$: 
 They reflect the internal structure of the  Cooper pairs
 in color superconductor. 
  These correlations  are related to the off-diagonal (1-2) element
  of the quark propagator:
\beq
 &&\langle\psi^a_i(t,\bfm{x})\psi^b_j(t,\bfm{y})\rangle_{\rm super}
   =\lim_{x^0\to y^0+\epsilon} i [S_{12}]_{ij}^{ab}(x-y)\nonumber\\
 &&=\int\frac{d^3\bfm{q}}{(2\pi)^3}%
    e^{i \scriptsize{\bfm{q}(\bfm{x}-\bfm{y})}}\left(%
 \frac{ \Lambda_+(\hat{\q} )\Delta_+(q) }{ 2\sqrt{E_+(q)^2+|\Delta_+(q)|^2}}%
 +\frac{ \Lambda_-(\hat{\q} )\Delta_-(q) }{ 2\sqrt{E_-(q)^2+|\Delta_-(q)|^2}}
          \right)(i\gamma_5C)(\lambda_2\tau_2)^{ab}_{ij}.
\eeq
$\hat{\varphi}_{\pm}(q)$ is simply extracted from the above and
 ${\varphi}_{\pm}(r)$ is defined as the Fourier transform
\beq
\hat{\varphi}_{\pm}(q)=  
\frac{\Delta_\pm(q)}{2\sqrt{E_\pm (q)^2
+|\Delta_\pm (q)|^2}}, \quad
\varphi_{\pm}(r)= N \int\frac{d^3\bfm{q}}{(2\pi)^3}
\  \hat{\varphi}_{\pm}(q)\  e^{i \bfm{q}\bfm{r}}, \label{correlation}
\eeq
\vspace{0.2cm}
where $N$ is a normalization constant determined by
$\int d^3r  |\varphi_+ (r) |^2 =1$.

\noindent
(iii) The coherence length
 $\xi_{\rm c}$  characterizing the 
 typical size of a Cooper pair:
  It is defined simply as a root mean square radius
 of $\varphi_{+}(r)$: 
\beq
  \xi_{\rm c}^2 
 = \frac{\int d^3r \  r^2 |\varphi_+ (r) |^2}{ \int d^3r 
   |\varphi_+ (r) |^2}
 =
\frac{\int_0^\infty dk\ k^2
             \left| d\hat{\varphi}_{+}(k)/ dk  \right|^2}{
 \int_0^\infty dk\ k^2\left|\hat{\varphi}_{+}(k)
    \right|^2 } .
  \label{coherence-length}
\eeq
A measure of the 
  coherence length $\xi_{\rm c}$ 
 in the weak-coupling limit is known as the Pippard length,
 which is given by $\xi_{\rm p}=(\pi\Delta_+(\mu))^{-1}$\cite{Fetter-Walecka}.
It is shown in Appendix B  that the quark correlation
  $\varphi_+(r)$ in the weak-coupling limit behaves as 
\beq
\varphi_+(r \rightarrow \infty )\propto  
 \frac{ \sin (\mu r)}{(\mu r) ^{3/2}} \cdot
  e^{-r/(\pi \xi_{\rm p})}.\label{asymptotic_WF}
\eeq

In a typical type-I superconductor  in metals, the Pippard length is of
  the semi-macroscopic order $\xi_{\rm p} \sim 10^{-4}$ cm, whereas
  inverse
  Fermi momentum is of the microscopic order $k_{\rm F}^{-1}\sim 10^{-8}$
  cm.
 The inverse of the Debye cutoff   is in between the two scales
   $\omega_{\rm D}^{-1} \sim 10^{-6}$ cm.
 Therefore there is a clear scale  hierarchy,
   $\Delta \ll \omega_{\rm D} \ll k_{\rm F}$.
 Because of  the absence of the intrinsic   scale $\omega_{\rm D}$,
  similar  scale hierarchy  in QCD at extremely high density reads 
    $\mu e^{-c/g} \ll \mu$.  
 At lower densities, however,  such scale separation becomes
    questionable for $g$ is not small.


\section{Numerical results}

 In this section, we present numerical results of the momentum-dependent
    gap and the other physical quantities.
 In Sect.~IIIA, we show solutions of the gap equations at very high
     density.
 Then we discuss whether the result has similarity
    to that in the BCS superconductivity for metals [see Eq.~(\ref{BCS})].
 What makes the color superconductor unique is the absence of
    the intrinsic cutoff scale  $\omega_{\rm D}$.
 Nevertheless, we will see that similar relation as Eq.~(\ref{BCS})
    holds under the replacement $\omega_{\rm D} \rightarrow \Delta $
    at least at extremely high density.
 We will also examine how the results of the gap equation in the
    weak-coupling limit (\ref{gap_simple1}) are modified when the effects
    such as the color-electric interaction, the momentum dependent
    coupling
    and the antiquark-pole contribution are taken into account.

 In Sects.~IIIB and IIIC, we repeat the same calculations
    at lower densities and show a qualitative difference from the
    weak-coupling limit.
 Substantial modification of the Fermi surface at low densities
    will be explicitly shown by computing the occupation number.
 Quark correlation in the color superconductor and  the size of the Cooper
    pair are calculated in Sect.~IIID.
 The results indicate that the color superconductivity at low
      densities
    is no longer similar to the usual BCS-type superconductivity.

 In Sects.~IIIA and IIIB, starting from the simplest gap equation
    in the weak-coupling limit (\ref{gap_simple1}),
    we will include the contributions of color-electric interaction,
    momentum-dependent coupling and antiquark pole,
    step by step.
 This procedure clarifies the importance of each contribution.
 Let us define each step below, for later convenience.

\begin{description}

\item[Step 1:]
        We solve the gap equation for $\Delta_+$ in the weak-coupling limit 
        Eq.~(\ref{gap_simple1}) where only the {\em color-magnetic}
        interaction is taken into account. The momentum {\em independent} 
        coupling $g^2(\mu,\mu)$ is used in this step.
\item[Step 2:]
        The Debye-screened {\em color-electric} interaction is further
        added to Step 1. This corresponds to solving 
        Eq.~(\ref{full-gap}) for $\Delta_+$ with  neglecting the 
        antiquark-pole contribution and with making a replacement 
        $g^2(q,k) \rightarrow g^2(\mu,\mu)$. 
\item[Step 3:]
        Same as Step 2 except for  
        the use of $g^2(q,k)$  instead of $g^2(\mu,\mu)$.
\item[Step 4:]
        The antiquark-pole contribution is also taken into account in Step 3.
        This gives the complete solution of our full coupled gap equations
        Eq.~(\ref{full-gap}).
\end{description}

Throughout the all steps, $g^2$ in the Debye screening mass $m_{\rm D}$ 
  is taken to be  $g^2(\mu,\mu)$ for simplicity. In Step 3 and Step 4, we 
  use a phenomenological value $\Lambda =400 {\rm MeV}$ and 
  $p_c^2=1.5\Lambda^2$ for $g(q,k)$ as we have already mentioned 
  in  Sect.~IIA.

\subsection{Momentum dependence of the gap at high density}

\begin{figure}[htb]
  \centerline{
  \epsfsize=0.8\textwidth
  \epsfbox{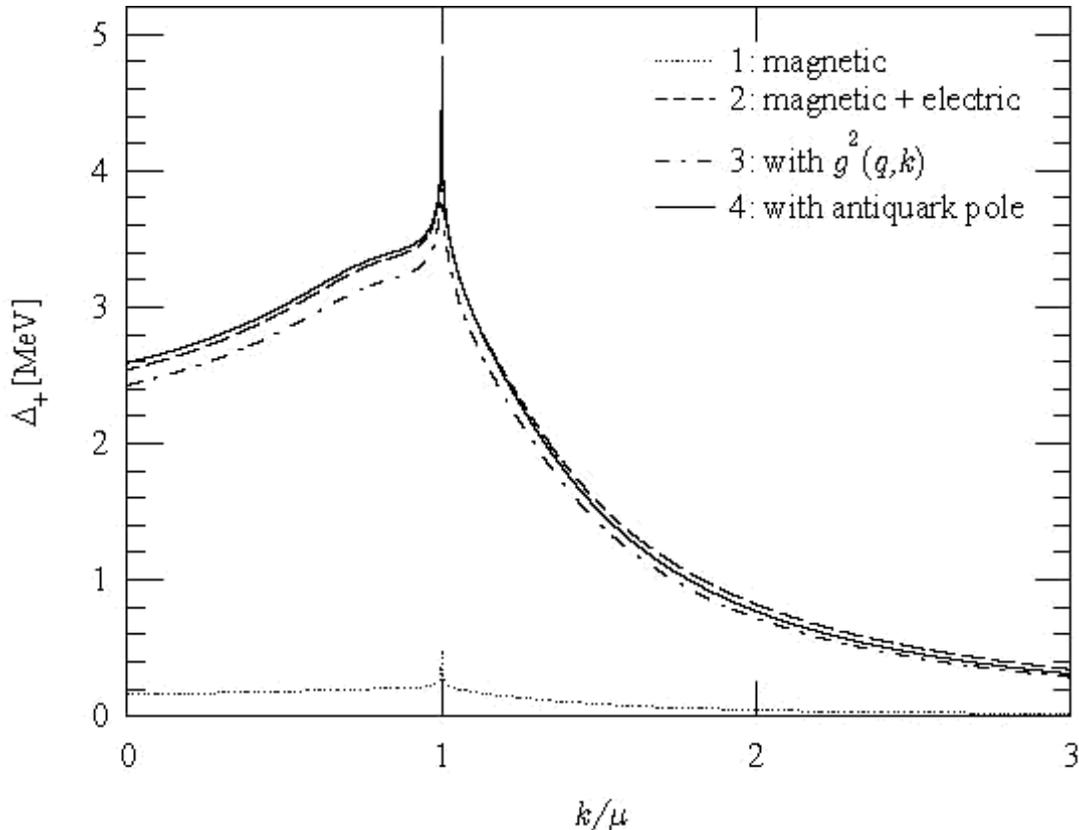}
  }
 \caption[]{The momentum dependence of the gap $\Delta_+$ from 
            Step 1 to Step 4 at high density, $\mu=2^{12}\Lambda$.
         The dotted line is a result of the magnetic interaction 
                 alone (Step1). 
         The dashed line is the gap with both magnetic 
                 and electric-interactions included (Step2). 
         The dot-dashed line is the gap in which the momentum dependent 
                  coupling $g^2(q,k)$ is used (Step 3) instead of the 
                  momentum  independent coupling $g^2(\mu,\mu)$ (Step 2).
         The solid line is the gap as a solution of the full gap 
                  equation (\ref{full-gap}) with $g^2(q,k)$ and 
                  the antiquark pole (Step 4).
}
 \label{fig:1}
\end{figure}

In this subsection,  we solve the gap equations at chemical potential
    $\mu=2^{12}\Lambda \simeq 1.6$TeV.
 This corresponds to the baryon density
    $\rho_{\rm B}=1.1\times 10^{11}\rho_0$ with
    $\rho_0=0.17{\rm fm}^{-3}$ being the normal nuclear matter density.
 At this extremely high density, we expect that the full gap equation
    Eq.~(\ref{full-gap}) is well approximated by its weak-coupling limit,
    Eq.~(\ref{gap_simple1}). Furthermore, the analytic solution 
    Eq.~(\ref{sol-LD}) is expected to give a 
     fair approximation of  Eq.~(\ref{gap_simple1})
    around the Fermi surface.
  In Fig.~\ref{fig:1}, we show the numerical results of the four steps.
     It is evident that all the results have a very narrow
     peak at the Fermi surface.
     This peak is not a singular one, but has a plateau with the width of 
     $2\Delta_+(k=\mu)$, as we will show later. 
     At high momentum $k\gg \mu$, the decrease of the gap obeys the power law.
     This is consistent with the analytic solution Eq.~(\ref{power}).
     Such a long tail is due to the unscreened nature of the color-magnetic
     interaction and is quite different from the behavior of the gap
     in the standard BCS-type superconductivity.
     Let us look into each step in more detail.\\

\noindent{\bf Steps 1 and 2:}

 Comparing the gap with the magnetic interaction alone 
    (the dotted line) and the gap with both magnetic and electric interactions
     (the dashed line), we find that 
   the electric interaction induces a large enhancement (about 10 times) 
   of the gap almost independently of momentum.
 This is already suggested by the analytic studies in Sect.~IIC. 

  This enhancement may be understood in a qualitative manner.
  In the coordinate space, the Debye-screened electric interaction behaves as 
  a Yukawa potential. Such a short-range interaction can form only a loosely 
  bound Cooper pairs with a large size.
  In fact, if one solves the gap equation with the electric interaction alone, 
   one finds  a very small gap compared to the one in Fig.~2(a). The situation 
   becomes different when the magnetic and electric interactions coexist: 
   Small size Cooper pairs are formed primarily by the long-range magnetic
   interaction. Then,  even the short-range electric interaction becomes
   effective to  generate further attraction between the quarks.
   This cooperative effect can qualitatively explain the reason why addition 
   of the electric interaction enhances the gap.\\
 

\noindent{\bf Steps 3 and 4:}

To see the effect of the momentum dependent coupling  $g^2(q,k)$,
 let us compare  the dot-dashed line  (where $g^2(q,k)$ 
 is used) with the dashed line (where $g^2(\mu,\mu)$ is used). 
Since $g^2(q,k)$ works as a weight factor in the momentum 
 integral in the gap equation,  substantial difference should appear between 
 the two cases if the contribution away from the Fermi surface is not 
 negligible in the integral. However, we find no significant difference 
 between the two cases at high density. 
 This implies that the color superconductivity at high density is governed 
 by the physics near the Fermi surface.

\begin{figure}[htb]
  \centerline{
  \epsfsize=0.49\textwidth
  \epsfbox{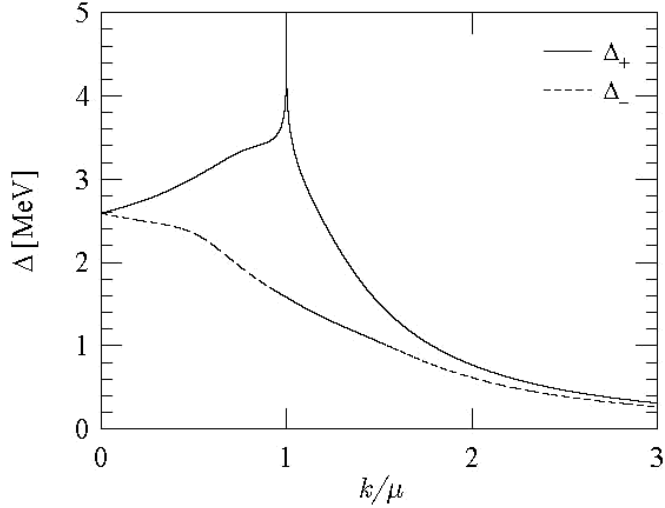}
  }
 \caption[]{$\Delta_+$ and $\Delta_-$ as the solutions to  
   the full coupled gap-equation in Step 4 at high density,
  $\mu=2^{12}\Lambda$. 
}
 \label{fig:2}
\end{figure}

Let us examine the last stage, Step 4, where the antiquark-pole
 contribution is included. 
 The solid line in Fig.~\ref{fig:1} is $\Delta_+(k)$
 obtained by solving the {\it full} coupled gap-equation (\ref{full-gap})
  for $\Delta_+$ and $\Delta_-$. 
 Since the difference from the previous step is small, 
 we can conclude that the quark-pole dominance is indeed a good approximation 
  at very high density.
The antiquark gap $\Delta_-(k)$ is shown in a separate figure 
  (Fig.~\ref{fig:2}) together with $\Delta_+(k)$ (the same as the 
   solid line in Fig.~1).  
 Unlike $\Delta_+(k)$, the antiquark gap $\Delta_-(k)$ is a smoothly 
 decreasing function of the momentum. Although $\Delta_-$ is not small 
 compared to $\Delta_+$, it  does not imply that the sizable antiquark 
 Cooper pairs exist 
 in the system because the number of antiquarks are much smaller than quarks 
 as we will show in Sect.~IIID.

\begin{figure}[htb]
    \centerline{
      \epsfxsize=0.49\textwidth
      \epsfbox{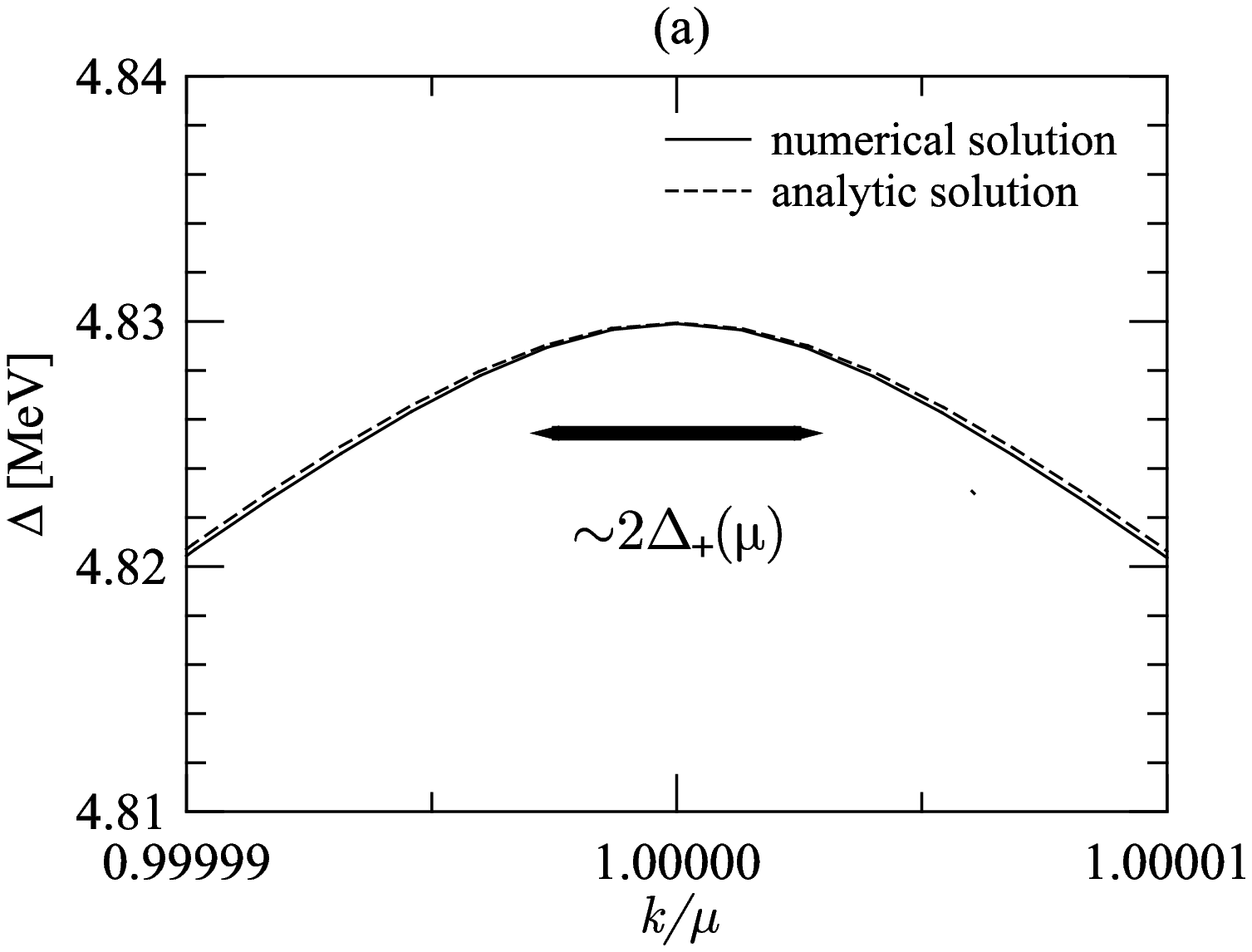}
      \epsfxsize=0.47\textwidth
      \epsfbox{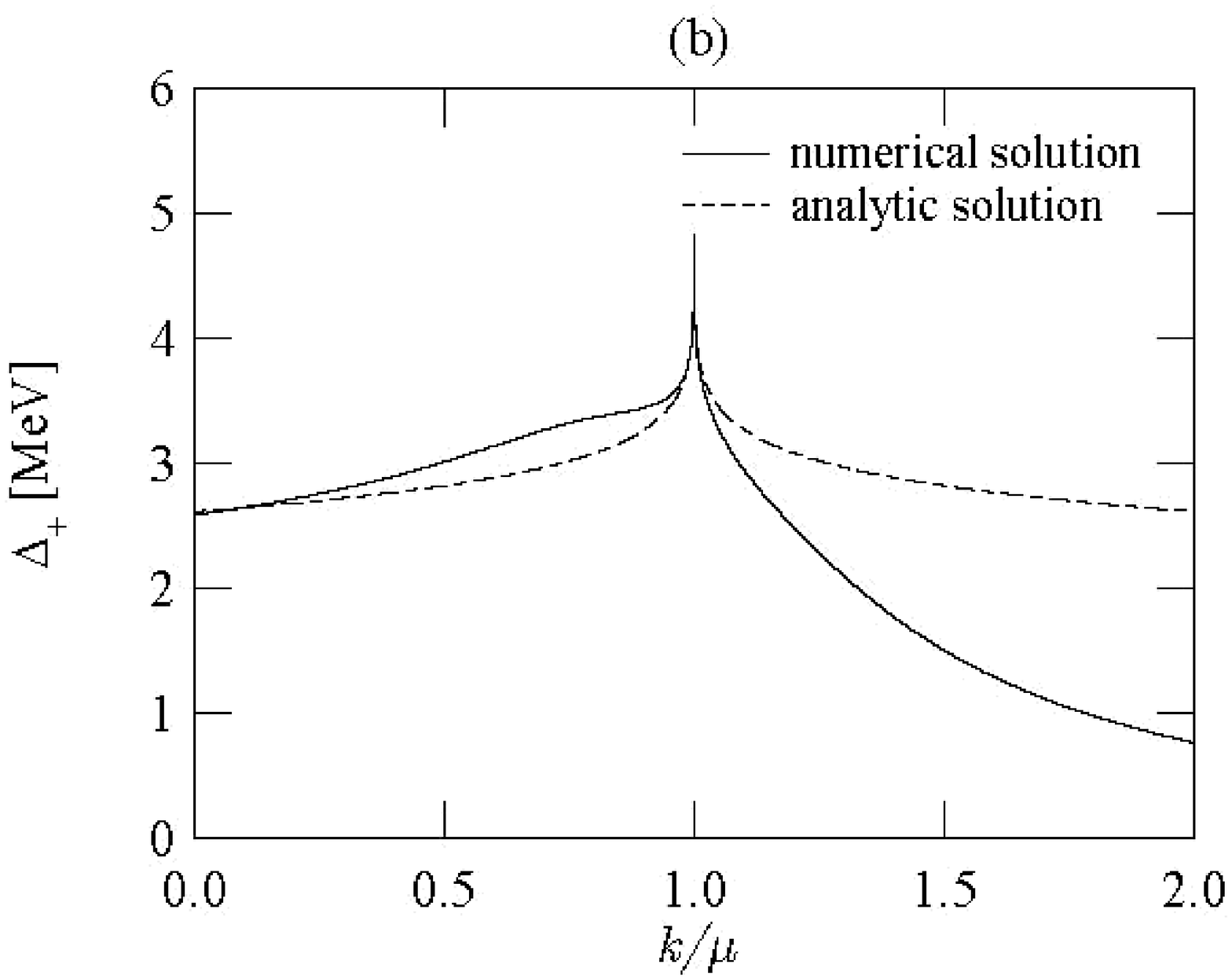}
    }
    \caption{Comparison of the numerical solution of $\Delta_+$
    in Step 4 (the solid line) and the analytic solution (\ref{sol-LD})
    (the dashed line) at high density, $\mu=2^{12}\Lambda$.
    The latter is normalized to the former at $k=\mu$.
    (a) near the Fermi surface, (b) global behavior.
              }
  \label{fig:3}
\end{figure}

 Finally, Fig.~\ref{fig:3} shows  a comparison between the momentum dependence
 of the numerical solution $\Delta_+$ in Step 4 (the solid line which is the
 same as the solid line in Fig.~\ref{fig:2}) with the analytic solution near
 the Fermi surface (\ref{sol-LD}) (the dashed line). The peak height of the 
 analytic solution is adjusted so that the solid and dashed lines coincide at
 $k = \mu $. In the vicinity of the Fermi surface (Fig.~3(a)), 
 the momentum dependence of the analytic solution agrees with the numerical 
 solution reasonably well. Note that  the peak is not a singular
 one as we mentioned before. On the other hand, as is shown in Fig.~3(b)
 the deviation becomes considerable away from the Fermi surface.
 In particular, the asymmetry with respect to the Fermi surface $k=\mu$ 
 represents the deviation from weak coupling analytic solution.
 Note that the coincidence at $k \sim 0$ is accidental because
 the analytic result  (\ref{sol-LD}) is valid  only near the Fermi surface.\\

 In this subsection,
  we have solved the momentum-dependent gap equations at high
  density $\mu=2^{12}\Lambda $. The characteristic features 
  of the momentum-dependent gap are 
  (i) there is a sharp peak at the Fermi surface,
      and 
  (ii) the gap decays rapidly but 
 is nonzero for momentum far away from the Fermi surface.
  The property (i) is similar to the standard BCS superconductivity
  [see Eq.~(\ref{BCS})] but (ii) is not, due to the absence of intrinsic
  ultraviolet  Debye-cutoff of the gluonic interaction in QCD.
  As for the magnitude of the gap at high density,
  the color-electric interaction
  enhances the gap considerably. The effects of the momentum-dependent 
  coupling $g^2(q,k)$ and the antiquark pairing are shown to be not 
  important at high density. 

\subsection{Momentum dependence of the gap at low density}

\begin{figure}[htbp]
  \centerline{
\epsfsize=0.8\textwidth
\epsfbox{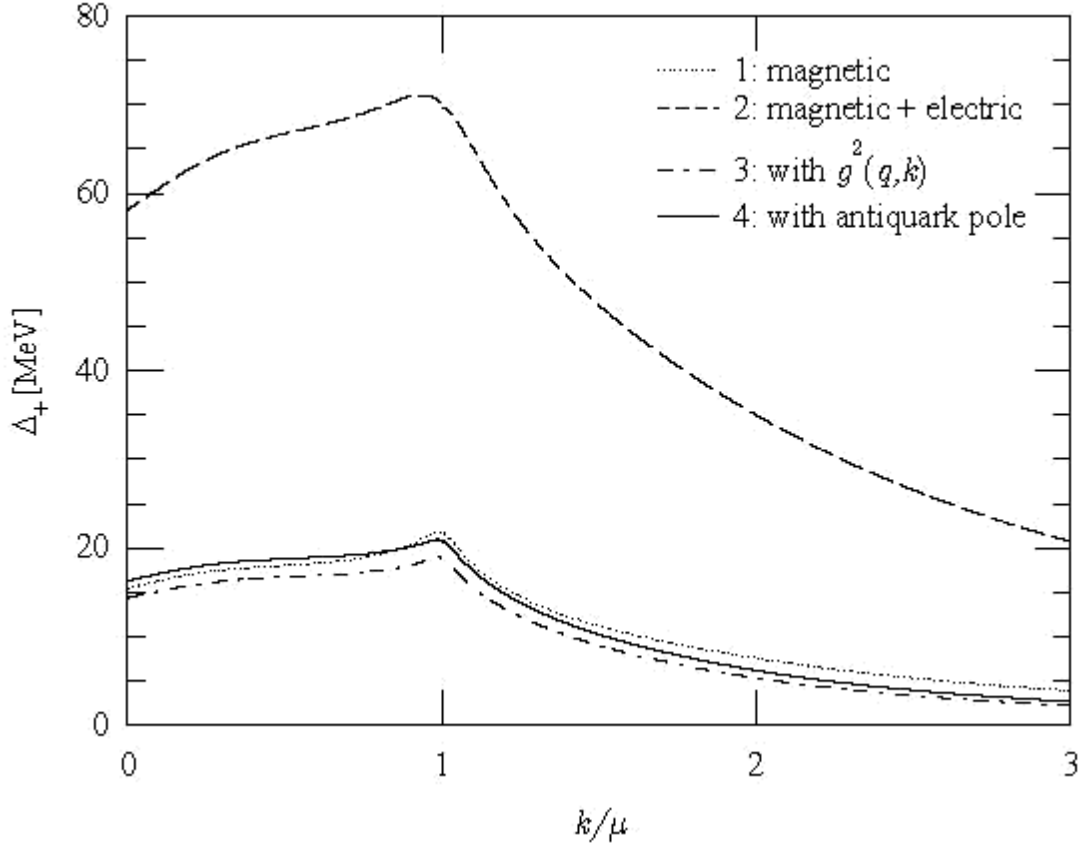}
  }
  \caption{Same as Fig.~\ref{fig:1} except that the  density is lower,
  $\mu=2\Lambda =800$MeV. }
  \label{fig:4}
\end{figure}

In this subsection, we carry out the same calculations as in Sect.~IIIA
    at lower density $\mu=2\Lambda=800$MeV (which corresponds
    to the baryon density $\rho_{\rm B}=13.2 \rho_0$). 
In Fig.~\ref{fig:4}, we show the numerical results of the four steps.
 One of the main differences from the high density case is  
  the absence of a sharp peak. 
 The width of the peak in Fig.~\ref{fig:4} in terms of $k/\mu$ 
    is much wider than that in  Fig.~\ref{fig:1}.
 This large modification of the gap function will be attributed to the 
    larger coupling at low density, because the Cooper pairs far from 
     the Fermi surface are easily formed.
 Therefore, this implies that the physics of superconductivity is 
    no longer limited on the Fermi surface.
 We confirm this below by more detailed analysis of each steps. \\

\noindent{\bf Steps 1 and 2:}

  Comparing the dotted line (with the magnetic interaction alone)
  and the dashed line (magnetic and electric interactions)
  in  Fig.~\ref{fig:4}, we again find that the addition of the 
  color-electric interaction induces a large enhancement of the gap. \\

\noindent{\bf Steps 3 and 4:} 

 The result of Step 3 (where $g^2(q,k)$ is used) is shown as the dot-dashed
    line, which should be compared with the previous step (the dashed line)
    where $g^2(\mu,\mu)$ is used.
 In contrast to the high density case in Fig.~\ref{fig:1}, the momentum 
    dependent coupling makes the magnitude of the gap considerably 
    smaller.
 This is one of the evidences that the momentum integration is not 
    any more dominated by the small region near the Fermi surface.
 The reason why the suppression of the gap takes place instead of the
    enhancement is understood as follows. Consider the gap at the Fermi 
 surface $\Delta_+(k=\mu)$. In the momentum integration of the gap equation 
 Eq.~(\ref{full-gap}) at $k=\mu$,
 $g^2(q,k=\mu) < g^2(\mu,\mu)$ for $ \mu < q$, 
    while $g^2(q,k=\mu) = g^2(\mu,\mu)$ for $ \mu > q$.
 Therefore, the momentum-dependent coupling always acts 
    to reduce the magnitude of the integral as compared to the 
    momentum-independent one. 

 Now, let us include the antiquark excitation in the gap equation (Step 4).
 The solution of our full gap equation (\ref{full-gap}) is shown 
   by the solid line. The antiquark-pole contribution 
    enhances the gap by 10\% .
 The difference between the solid and dot-dashed lines is 
    not huge but the absolute magnitude of the enhancement is larger
    than that in the extremely high density case shown  in Fig.~1.
 The reason for this enhancement is clear: As one decreases density,
   the chemical potential becomes small, which plays a role of the 
   intrinsic energy gap for antiquark excitation.
 Therefore, the antiquark-pole contribution becomes non-negligible.
 This also suggests that the more low-momentum antiquarks are present 
   at low density which will be confirmed in Sect.~IIID by computing 
   the occupation number of antiquarks.
 
\begin{figure}[htbp]
  \centerline{
\epsfsize=0.49\textwidth
\epsfbox{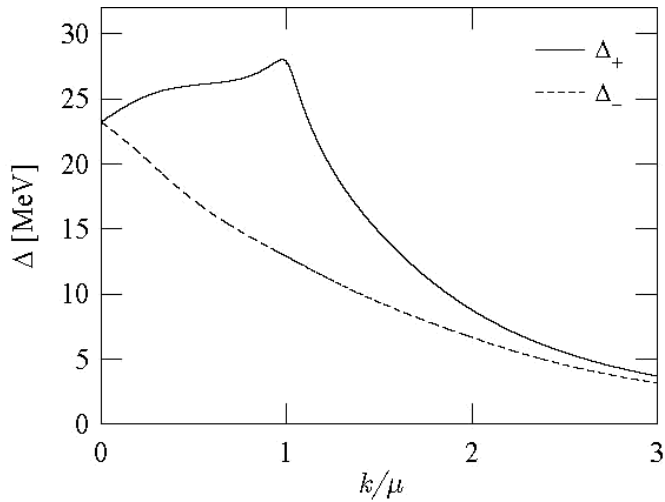}
  }
  \caption{Same as Fig.~\ref{fig:2} except that the  density is lower,
$\mu=2\Lambda =800$MeV. }
  \label{fig:5}
\end{figure}

 Finally, in Fig.~\ref{fig:5}, we compare the quark-gap $\Delta_+$ (the 
 solid line) and the antiquark gap $\Delta_-$ (the dashed line).
 As in the case of high density, the antiquark gap $\Delta_-$ is
 a smooth function of $k$ and is  not small compared to $\Delta_+$.\\

 In this subsection, we found that the momentum dependence of the gap at
 low density is quite different from that at very high density.
 The sharp peak at the Fermi surface disappears.
 The gap equation in the weak-coupling limit (\ref{gap_simple1}) is
 no longer a good approximation and all the contributions neglected
 in the weak-coupling limit are not negligible.
 All these results allow us to conclude that the color superconductivity
 at low density  is not a phenomenon just around the Fermi surface.
 In subsections IIID and IIIE below,  we will strengthen this
 picture by looking at other quantities such as the quark occupation numbers 
 and the size of the Cooper pair.


\subsection{Density dependence of the gap}

\begin{figure}[htbp]
  \centerline{
    \epsfsize=0.49\textwidth
    \epsfbox{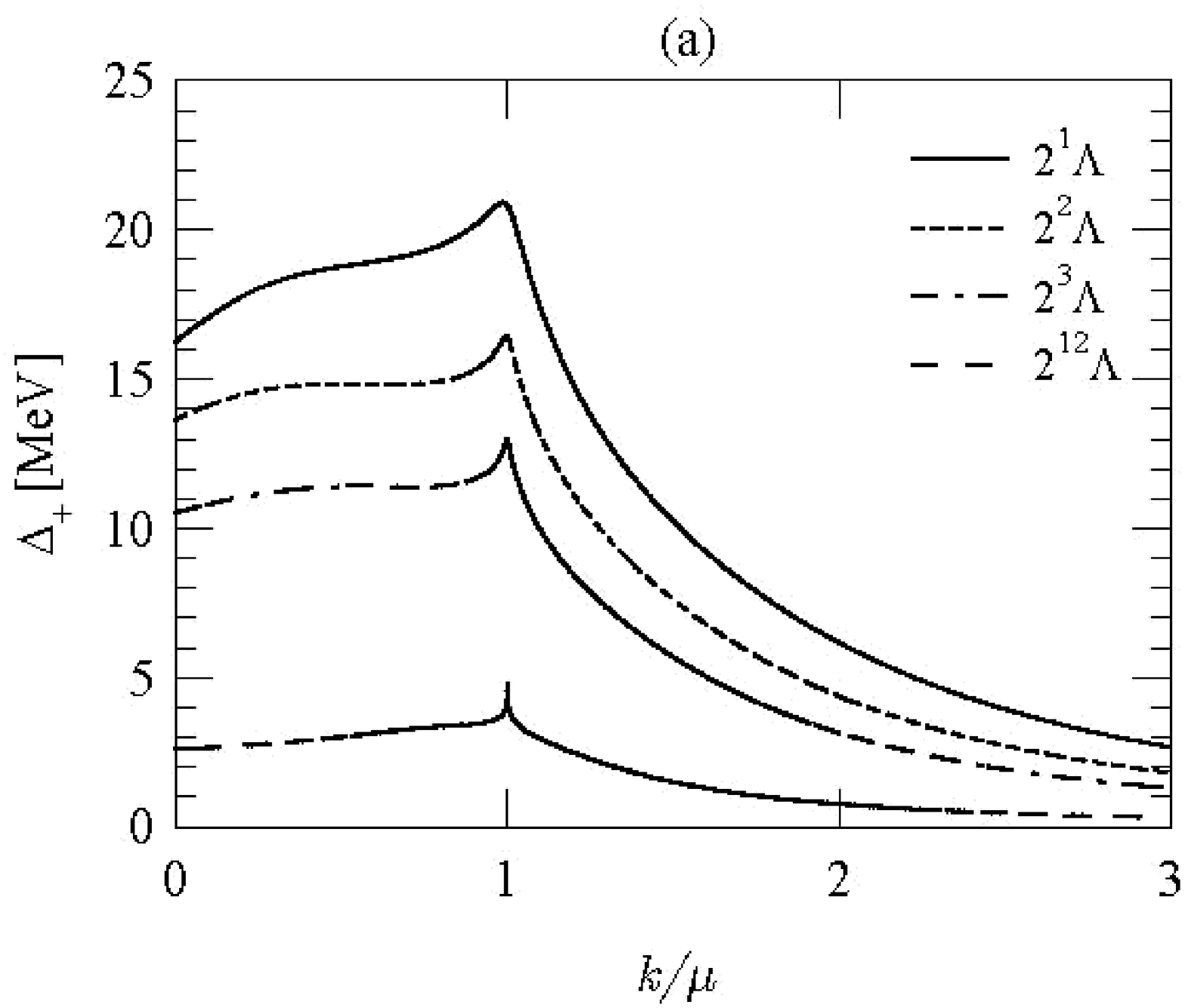}
    \epsfsize=0.49\textwidth
    \epsfbox{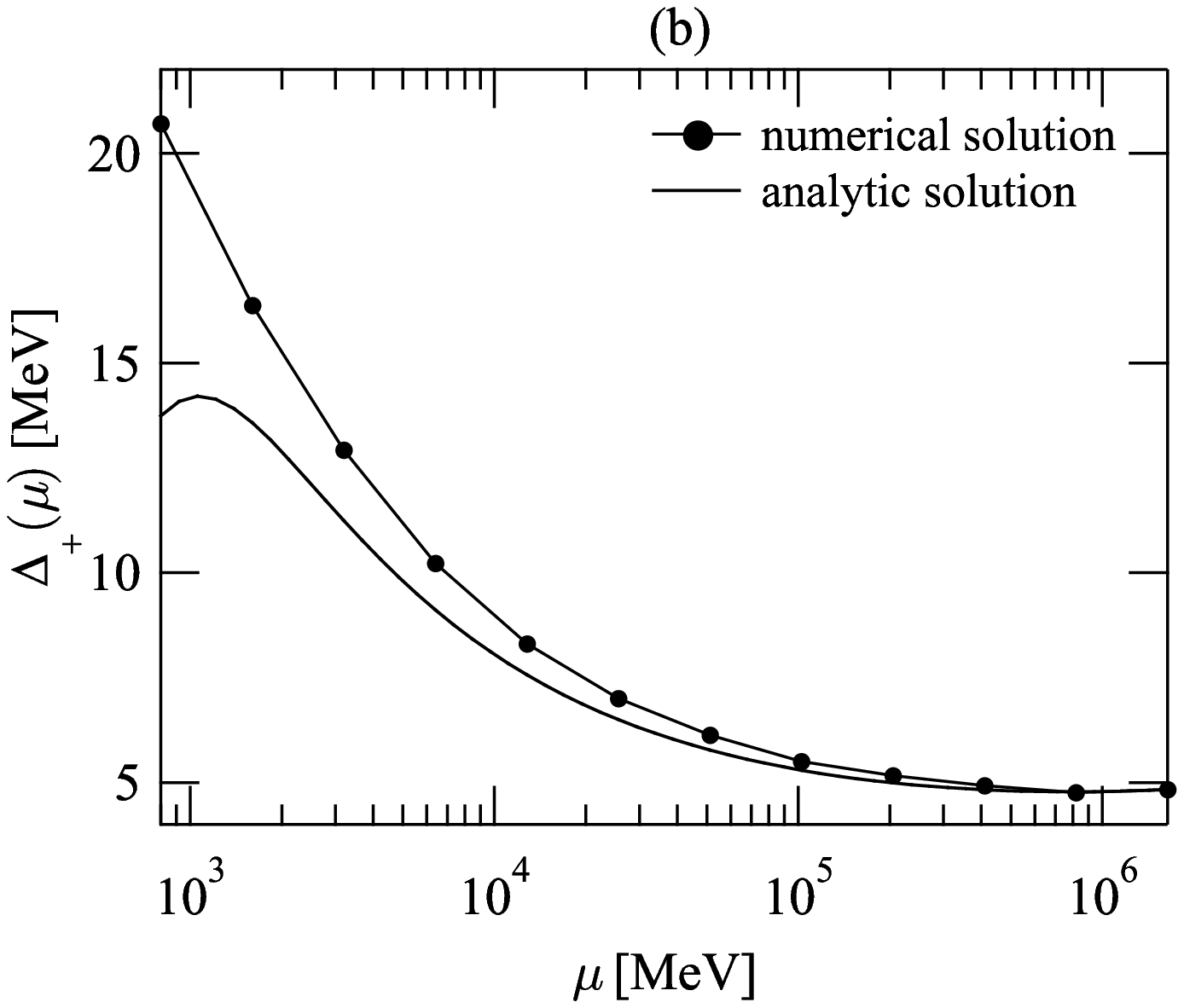}
  }
  \caption{%
    (a) $\Delta_+(k)$ as a function of $k/\mu$ for various densities 
        $\mu =2^n\Lambda $ with $n=1,2,3,12$. 
        All the calculations are done with the momentum dependent
        vertex $g^2(q,k)$ and with the antiquark-pole contribution.
    (b) Chemical potential dependence of the gap $\Delta_+(k=\mu)$  
        in the full calculation compared with the analytic result 
        which is normalized at the highest density $\mu = 2^{12} \Lambda$.}
  \label{fig:6}
\end{figure}

Let us now examine how the gap at high density  with a sharp peak 
	changes into the gap at  low density  with only a broad bump.
 In Fig.~\ref{fig:6}(a), we show the gap $\Delta_+(k)$ as a solution of
   the full gap equation (Step 4) for a wide range of densities.
  Since the actual position of the Fermi surface moves
   as we vary the density,  we use  $k/\mu$ as a horizontal axis in the figure 
    in order to show the change of the global behavior.
  The figure shows that the sharp peak at high density gradually gets
   broadened and simultaneously 
    the magnitude of the gap increases as we decrease the density.

 In Fig.~\ref{fig:6}(b),  the gap at the Fermi surface $\Delta_+(\mu)$ 
  is shown as a function of the chemical potential.  
  It decreases monotonically as $\mu$ 
  increases, but turns into an increase for $\mu > 10^6 $ MeV.
  The analytic solution is also shown 
  in Fig.~\ref{fig:6}(b). The magnitude of the analytic solution is normalized 
  to the numerical solution at the highest 
  density $\mu=2^{12}\Lambda \simeq 1.6 \times 10^6 {\rm MeV}$.
 At high density, 
  $\mu$-dependence of  the numerical result is in good agreement with the 
  analytic form which has  a parametric dependence 
  $\Delta_+(\mu) \propto  g^{-5}\mu\exp(-3\pi^2/\sqrt{2}g)$ with  
$g^2= g^2(\mu,\mu)$.
 On the other hand, the difference of the two curves at low density implies
 the 
  failure of the weak-coupling approximation. 
 As we have seen before, the use of $g^2(q,k)$ and the antiquark-pole has 
  non-negligible effects on the gap in the low density regime.

\subsection{Occupation numbers}

 So far we have seen that the weak-coupling picture
   of the color superconductivity is modified
   at low densities.
 In order to see how the Fermi surface is diffused by the Cooper pairing, 
    let us evaluate the occupation numbers of quarks and antiquarks.
    Since we have solved the full gap equation  
    (\ref{full-gap}) for  $\Delta_\pm(k)$, we can immediately obtain  
    the occupation numbers using Eq.~(\ref{occupation}).

\begin{figure}[htb]
 \vspace{2mm}
 \begin{minipage}{0.49\textwidth}
  \epsfsize=0.99\textwidth
  \epsfbox{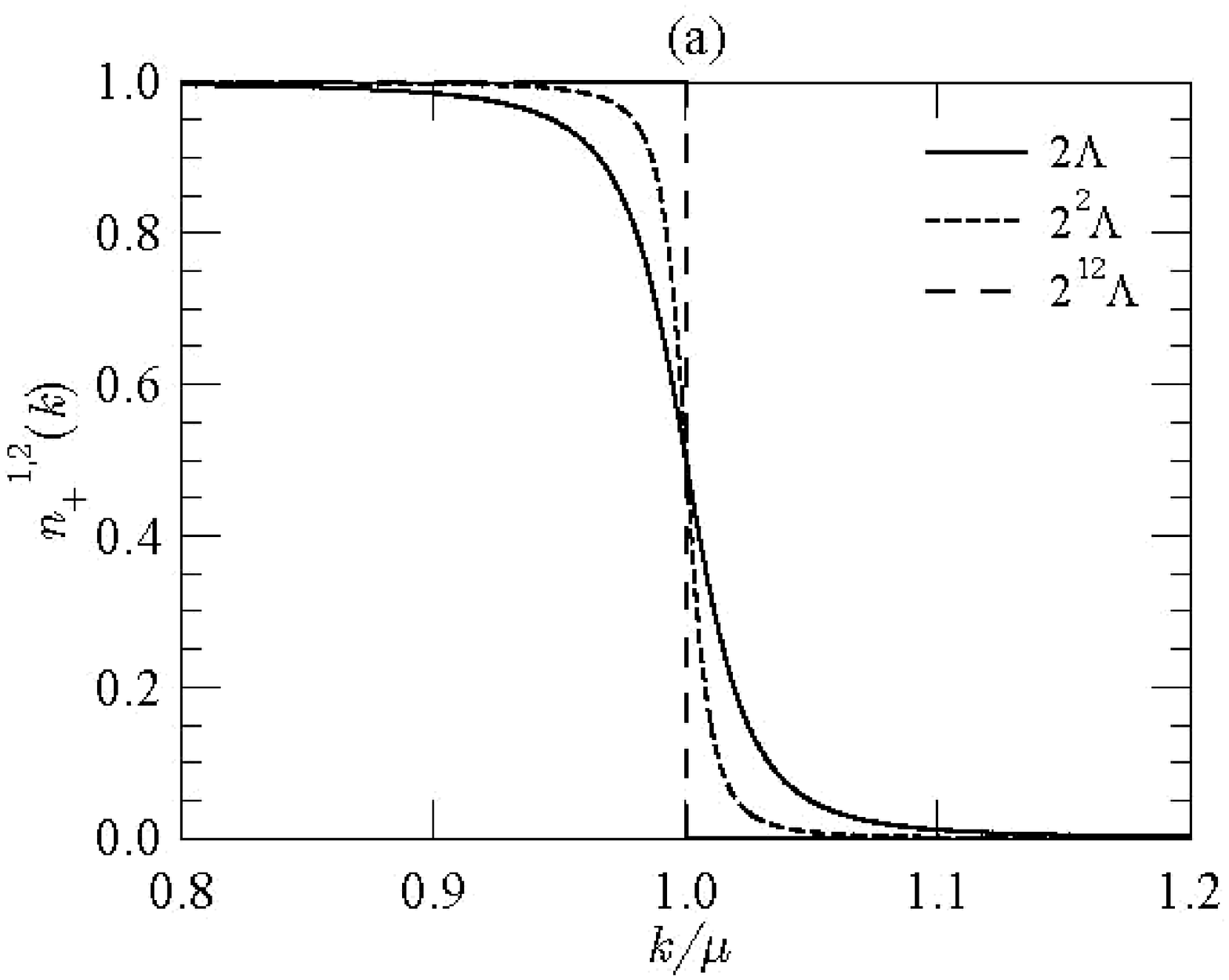}
 \end{minipage}%
 \hfill%
 \begin{minipage}{0.49\textwidth}
  \epsfsize=0.99\textwidth
  \epsfbox{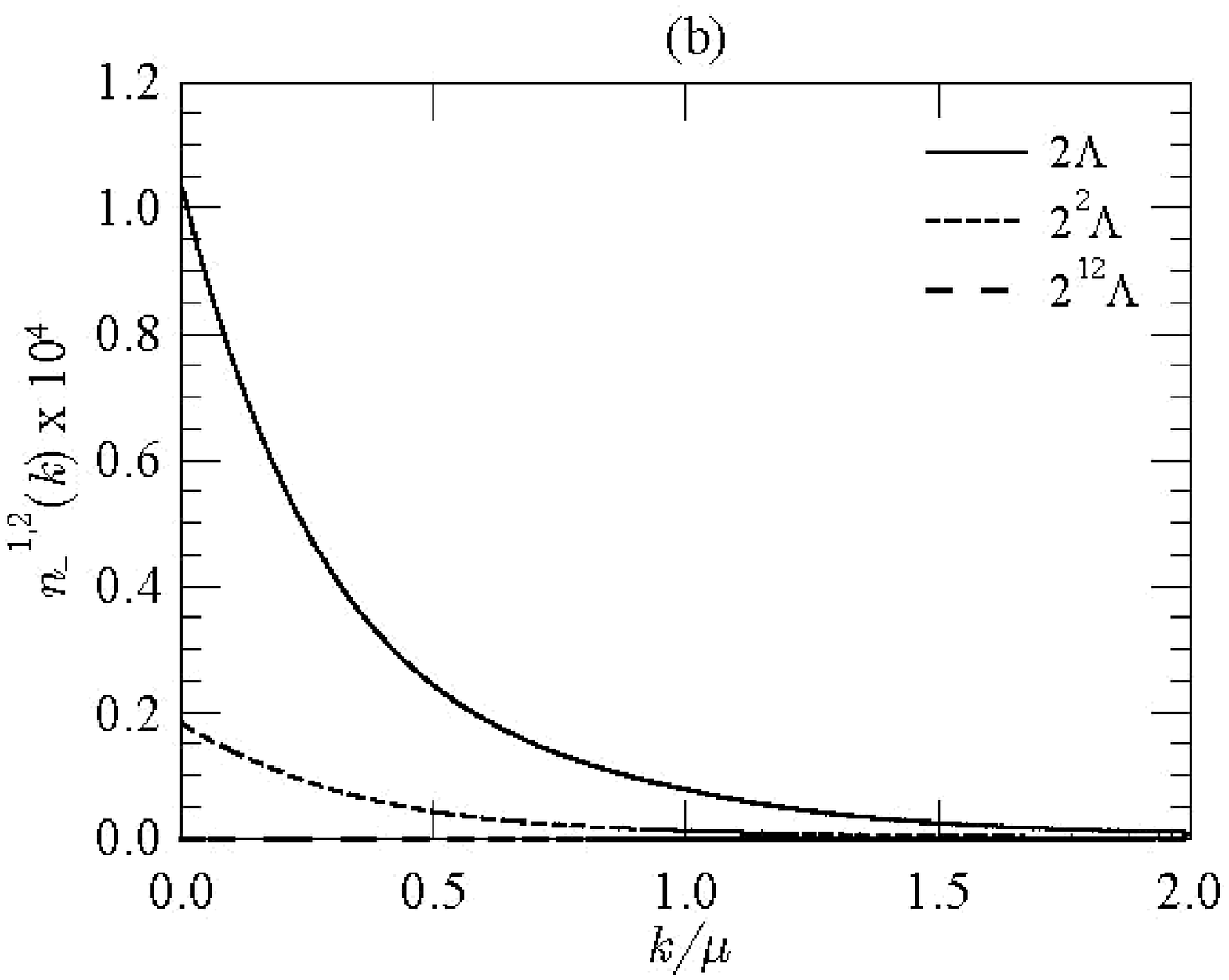}
 \end{minipage}%
 \caption{Occupation numbers of quarks $n_+^{1,2}$ (a) and
  antiquarks $n_-^{1,2}$ (b) for several  different
          densities.}
 \label{occupation_numbers}
\end{figure}

   In Fig.~\ref{occupation_numbers}, the occupation numbers of quarks 
   [Fig.~\ref{occupation_numbers}(a)] and antiquarks 
   [Fig.~\ref{occupation_numbers}(b)]     are shown at high density   $\mu=2^{12}\Lambda$ and at lower  densities $\mu=2\Lambda, 2^2\Lambda$.
   One finds that  the quark occupation numbers 
  is almost a step function at high density, while it is
    smeared out      for a wide region of momentum at low  densities. 
    The diffuseness of the Fermi surface
     is found to be of the order of $\Delta_+(\mu)$, which is consistent with
     the definition in  Eq.~(\ref{occupation}).

 Figure~\ref{occupation_numbers}(b)  
  implies that a small amount of the antiquarks also participate in the color
    superconductivity. Although the  antiquark gap $\Delta_-$ is 
    of the same order of the quark gap $\Delta_+$,
    the antiquark occupation number is generally suppressed
    due to the large
    energy denominator $\{(k+\mu)^2+|\Delta_-|^2\}^{1/2}$.
    As one decreases the density, however, such suppression is relatively
      weakened and  
       the magnitude of the antiquark occupation number increases,
        as can be seen from  Figure~\ref{occupation_numbers}(b).

\subsection{Correlation function and coherence length}

  One of the advantages  of treating the momentum dependent gap is that 
    we are able to calculate the correlation function which physically
    corresponds to the ``wavefunction'' of the Cooper pair. Such
    correlations  have been first
     studied in Ref.~\cite{Matsuzaki} in the 
     context of the color superconductivity.
     
\begin{figure}[htb]
 \vspace{2mm}
 \begin{minipage}{0.49\textwidth}
  \epsfsize=0.99\textwidth
  \epsfbox{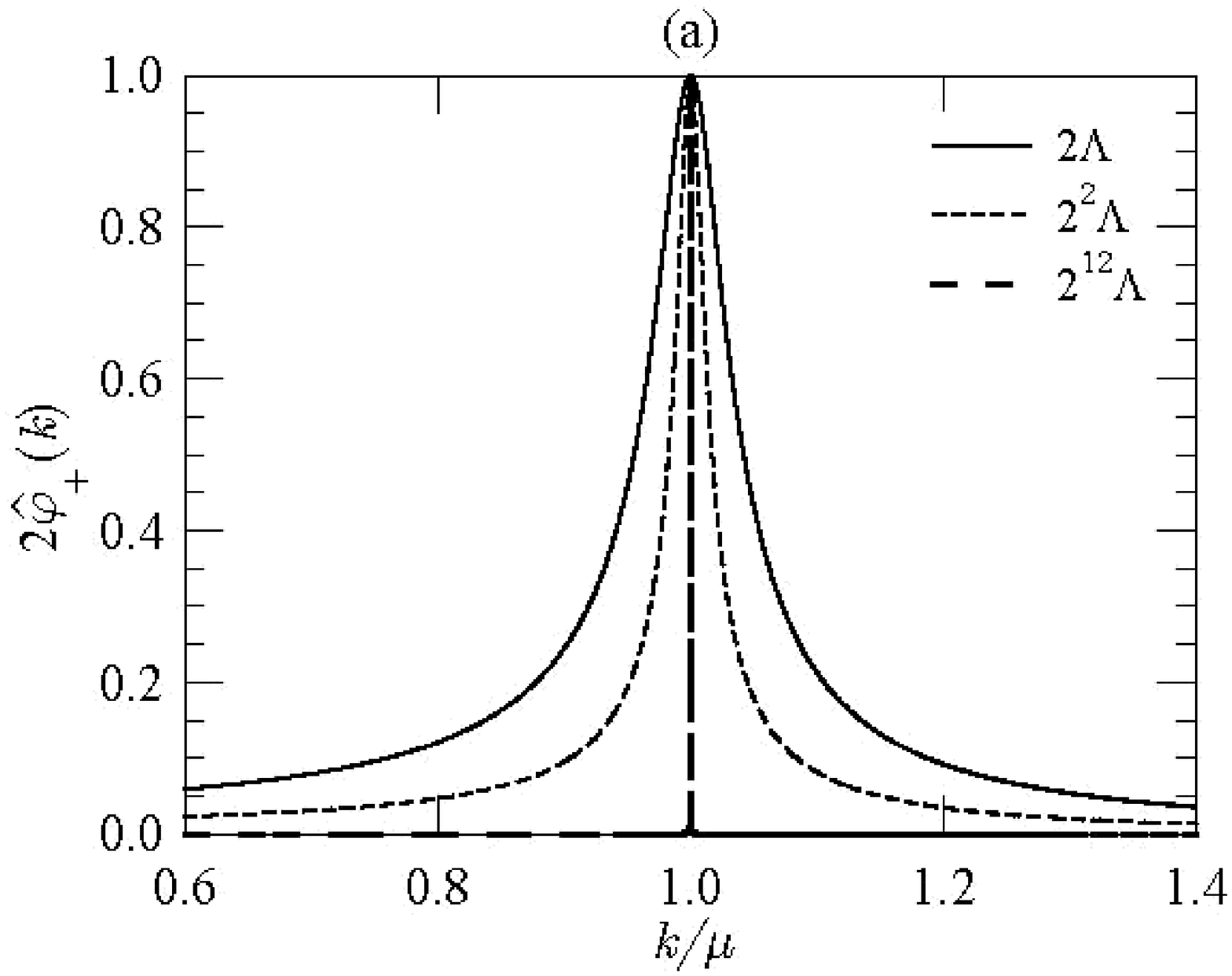}
 \end{minipage}%
 \hfill%
 \begin{minipage}{0.49\textwidth}
  \epsfsize=0.99\textwidth
  \epsfbox{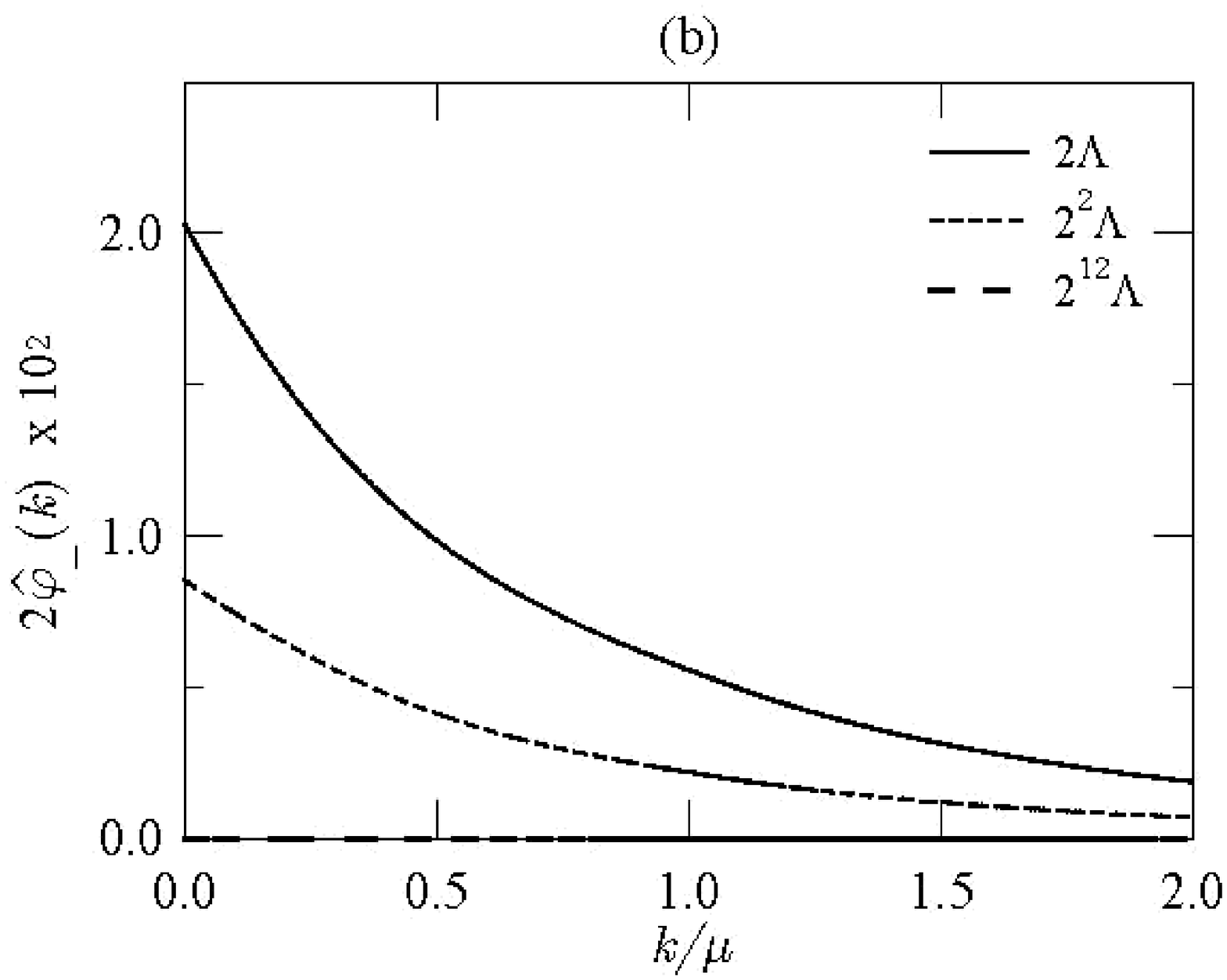}
 \end{minipage}%
 \caption{The correlation functions in the momentum space 
  at  several different densities 
         for quarks (a)  and for antiquarks (b). }
 \label{fig:Cor}
\end{figure}

       Using Eq.~(\ref{correlation}), we calculate the correlation functions 
    of quarks and antiquarks, $\hat\varphi_\pm(k)$.
  The results  at various densities are shown in Fig.~\ref{fig:Cor}.
  For quarks [Fig.~\ref{fig:Cor}(a)], the correlation function at very high 
    density has a sharp peak at the Fermi surface but it becomes
    broader as we decrease the density. 
  This is of course due to the broadening of the gap which 
  we found in Fig.~\ref{fig:6}.
 
  For antiquarks [Fig.~\ref{fig:Cor}(b)],  the correlation
  is much weaker than that of quarks and is a
  smoothly decreasing function of $k$. 
    Also, the magnitude of the  correlation increases as we decrease 
    the density. Since  $\Delta_-(k) \ll \mu$ holds  for densities
   considered in this paper, the above features can be
    simply understood by an approximate relation,
   $\hat{\varphi}_-(k)\sim\Delta_-(k)/2(k + \mu) $.
 
\begin{figure}[htbp]
 \vspace{2mm}
 \begin{minipage}{0.49\textwidth}
  \epsfsize=0.99\textwidth
  \epsfbox{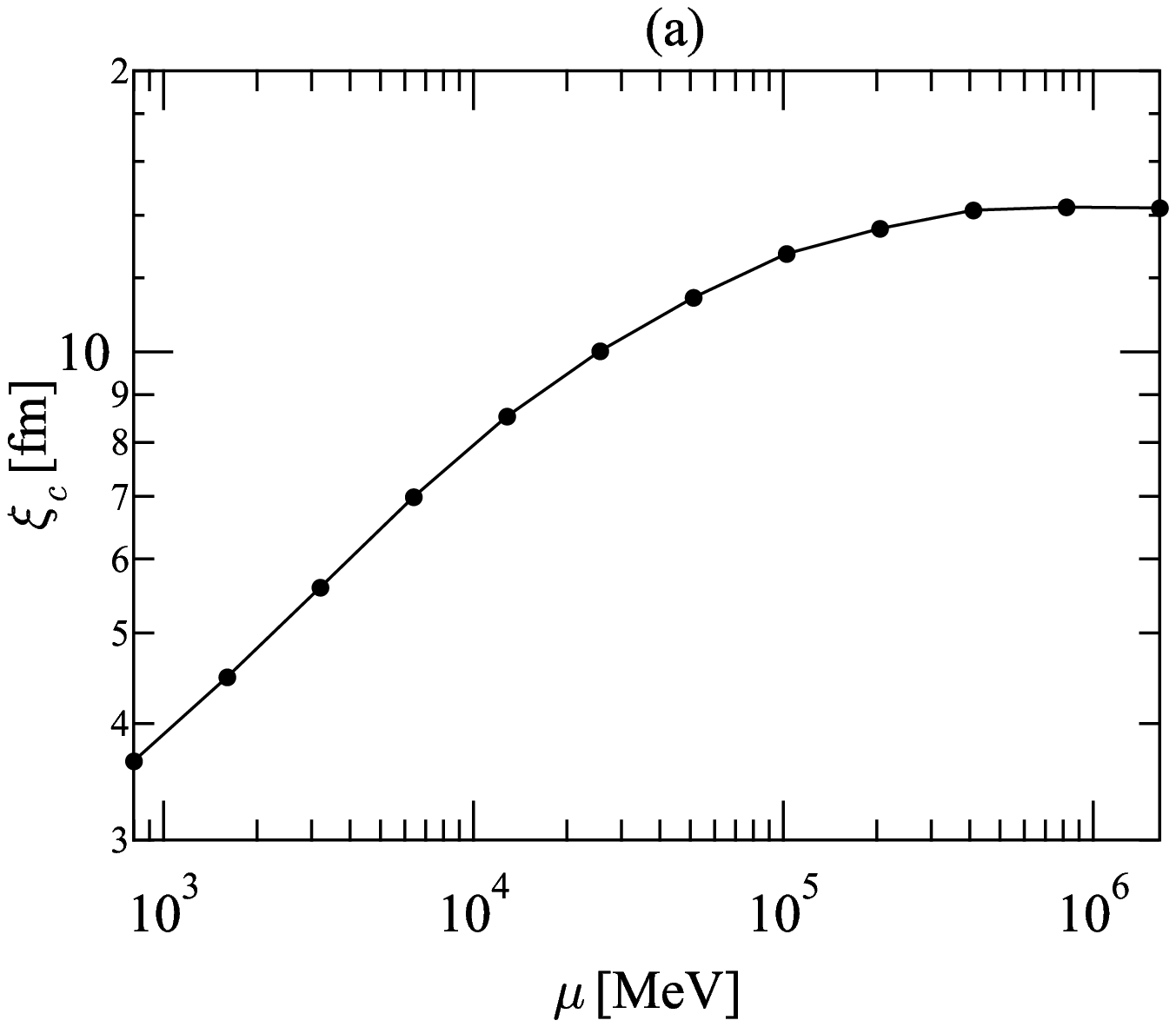}
 \end{minipage}%
 \hfill%
 \begin{minipage}{0.49\textwidth}
  \epsfsize=0.99\textwidth
  \epsfbox{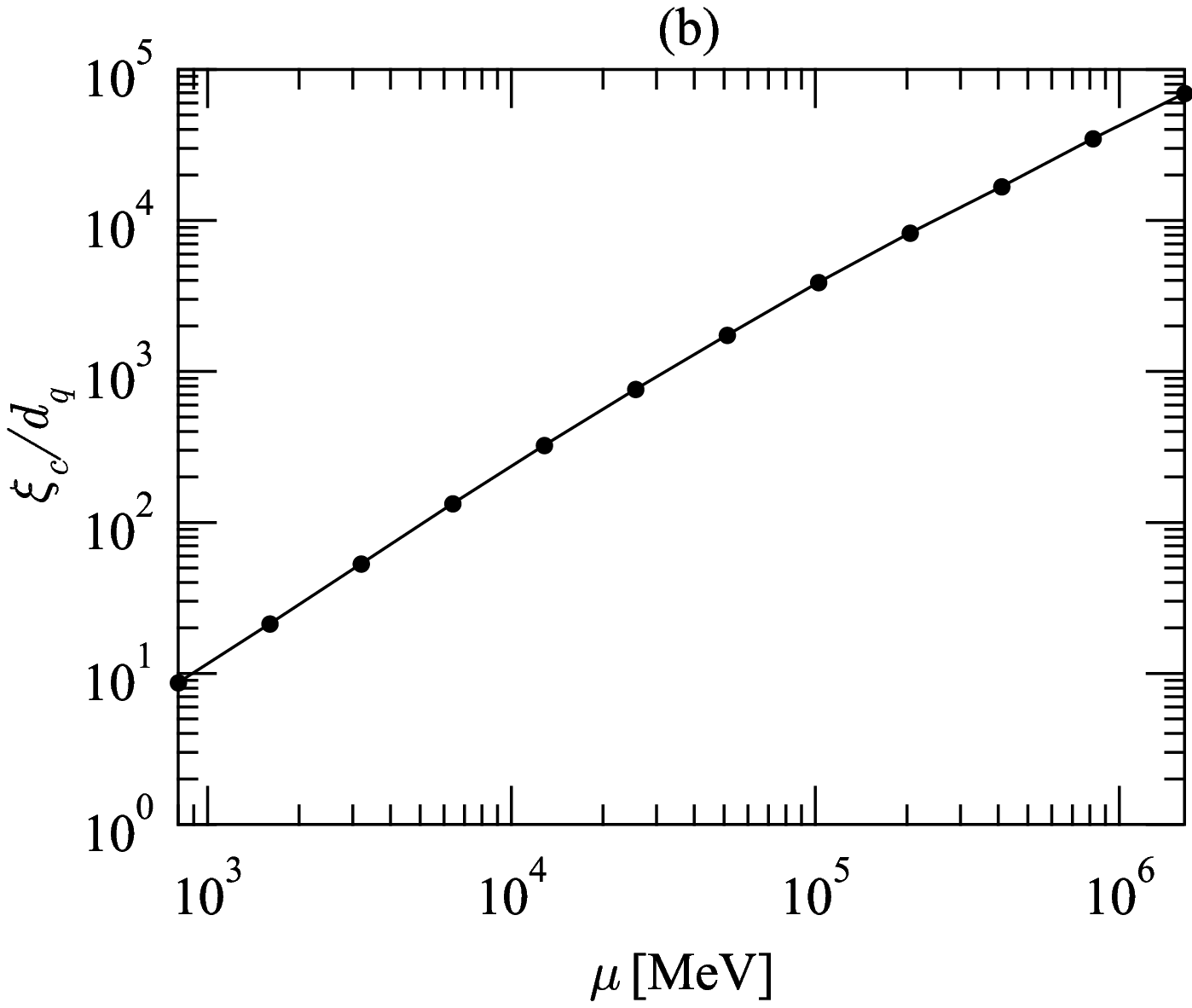}
 \end{minipage}%
  \caption{(a): Density dependence of the coherence length.
    (b): Ratio of the coherence length $\xi_{\rm c}$  and the
          average inter quark distance $d_q$ as a function of the
           chemical potential.}
  \label{Coh}
\end{figure}

 We can compute the size of a Cooper pair from the correlation function.
 Figure~\ref{Coh}(a) shows the coherence length $\xi_c$ of a quark Cooper pair 
    defined as the root mean square radius of the correlation function
    [see Eq.~(\ref{coherence-length})].
 The size of a Cooper pair becomes smaller as we go to lower     densities. 
 This  tendency   is understood by the 
 behavior of the  Pippard length $\xi_{\rm p}=1/\pi\Delta_+(\mu)$ 
  (which gives a rough estimate of the coherence length)
   together with the behavior of $\Delta_+(\mu)$ shown in 
 Fig.~\ref{fig:6}(b).

 Also,  the Cooper pair becomes 
   smaller  as density increases beyond  $\mu=2^{12}\Lambda$.
  However, it does not necessarily imply the existence of tightly bound 
    Cooper pairs. 
 In fact,
 the size of a Cooper pair  makes sense only   in comparison to the
  typical length scale of the system, namely
 the  averaged inter-quark distance $d_q$ defined  as
 \footnote{This is a result of 
    free quarks. To obtain accurate density $\rho$ and $d_q$, we have to 
    include contributions from interaction. However, such  correction is 
    suppressed by the power of $\Delta_+/\mu$.} 
$$
  d_q=\left(\frac{\pi^2}{2}\right)^{1/3}\frac{1}{\mu}.
$$
As we go to higher densities,   the ratio $\xi_{\rm c}/d_q$ 
 increases monotonically as shown   in 
    Fig.~\ref{Coh}(b). Namely  loosely bound Cooper pairs similar
      to the BCS superconductivity
    in metals are formed at extremely high densities.
 
 At the lowest density in Fig.~\ref{Coh}, 
 the size of the Cooper pair is less than 4 fm and 
  the  ratio $\xi_{\rm c}/d_q$ is less than $10$.
 The transition from  $\xi_{\rm c}/d_q \gg 1 $ to $\xi_{\rm c}/d_q \sim 1$
    as $\mu$ decreases is analogous to the transition from the BCS-type
     superconductor to the 
    so-called ``strong coupling'' superconductor.
  The weak-coupling BCS superconductivity
    may smoothly change  into the Bose-Einstein condensation (BEC) of 
    tightly bound Cooper pairs as the coupling strength increases%
    \cite{BEC}.
    Our result here suggests that the quark matter possibly realized in
    the core of neutron stars may be rather like the BEC of tightly bound
    Cooper pairs.

\begin{figure}[htb]
  \centerline{
\epsfsize=0.49\textwidth
\epsfbox{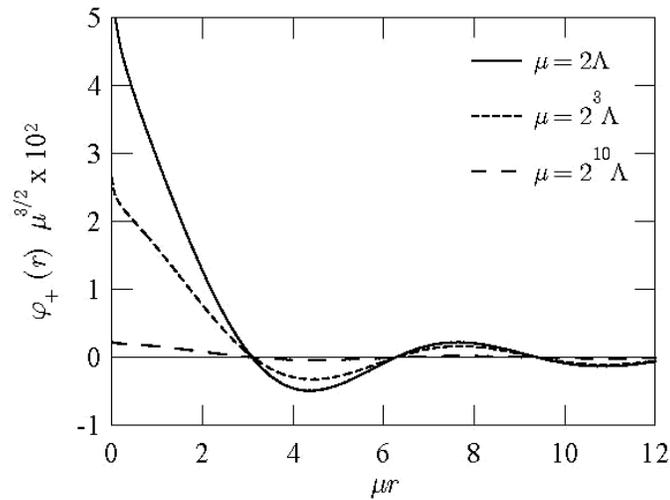}
  }
  \caption{The quark-quark  correlation function $\varphi_+(r)$ 
  in the coordinate space  $\varphi_+(r)$  for  
  several different  chemical potentials. $\varphi_+(r)$ is normalized to
   be unity.
 }
 \label{Spatial_WF}
\end{figure}

For better understanding of the internal structure of the quark Cooper pair,
  let us consider the correlation function in the coordinate space.
 Fig.~\ref{Spatial_WF} shows the spatial correlation 
    of a Cooper pair at various chemical potentials 
  normalized as
$
    \int d^3{r}|\varphi_+(r)|^2=1.
$
 As is expected, the density dependence of the quark correlation in
   the coordinate space is opposite to that in the momentum space. 
 At high density, most of the quarks participating in forming  
   a Cooper pair have the Fermi momentum  
   $k_{\rm F}=\mu$ giving a sharp peak in the momentum space correlation.
 In the coordinate space, this corresponds to an oscillatory
   distribution   with 
  a wavelength $\lambda=1/\mu$ without much structure
   near the origin. (The oscillation  is also evident from the
   factor $\sin (\mu r)$ in the 
  approximate correlation function Eq.~(\ref{asymptotic_WF}) discussed
   in Sect.~IIE.)  At lower densities, accumulation of the 
    correlation near the origin in the coordinate space
     is much more prominent in Fig.~\ref{Spatial_WF}. This  
 implies a localized Cooper pair composed of quarks with various
  momentum.  This is also seen
   by  the broad  momentum correlation in Fig.~\ref{fig:Cor}(a).

\section{Summary and Discussion}

 In this paper, we have studied the spatial-momentum dependence of a 
    superconducting gap and the structure of the Cooper pairs in 
    two-flavor color superconductivity, using a single model for a very 
    wide region of density.
 Nontrivial momentum dependence of the gap manifests itself at low 
    densities, where relatively large QCD coupling  allows the Cooper
   pairing to take place in a wide region around the Fermi surface.
 Our results imply that the quark matter which might exist in the core
    of neutron stars or in the quark stars could be  rather different 
    from that expected from the weak-coupling BCS picture.

Following is the summary of what we have discussed in this paper.

\begin{enumerate}
\item [(1)] 
  At high density, the weak-coupling gap equation with the electric and 
    magnetic gluons is a good approximation. 
  The momentum dependent quark-gluon vertex and the coupling to the antiquark 
    pairing do not change the weak-coupling result.
  The gap $\Delta_+(k)$ has a peak around the Fermi surface
    with a width $\sim2\Delta_+(\mu)$ and decreases rapidly as $k$
    goes away from the Fermi surface.
  This is consistent with the analytic solution of the gap equation
    in the weak-coupling limit. 
  All the results indicate that the Cooper pairing takes place only 
    in a small region around the Fermi surface.

\item[(2)] 
  At lower density, the weak-coupling gap equation is no longer a good 
    approximation.
  The momentum-dependent vertex and the coupling to the antiquark pairing have 
    non-negligible effects. 
  The sharp peak at the Fermi surface disappears.
  These imply that a large number of Cooper pairs away from the Fermi surface
    participate in the color superconductivity.
  This was confirmed by the quark occupation numbers and the quark-quark 
    correlation.
  Therefore, color superconductivity at low density is not a phenomenon 
    in the vicinity of the Fermi surface, but is a phenomenon with
    large modification of the Fermi surface.

\item[(3)] 
  The qualitative change of color superconductivity from high density 
    to low density can be explicitly seen by the ratio of the size of the 
    quark Cooper pair to the averaged inter-quark distance. 
  At high density, the ratio is very large (about $10^5$ at $\mu\sim 10^6$MeV) 
    which is consistent with the standard BCS-type superconductivity.
  At lower densities, however, the ratio becomes small (about $10$ at $\mu=800$MeV).  
  This situation is rather similar to the ``strong coupling'' superconductor 
    which could be described by a Bose-Einstein condensate of tightly bound
    Cooper pairs.

\end{enumerate}

There are several future problems.

Firstly, there are still several corrections to our ``full'' gap equation.
   They include the full hard-dense-loop corrections and the Meissner effect 
   in  the gluon propagator, and  also  the use of the plasmino dispersion 
   in the diagonal self-energy of the quark propagator. The latter effect 
   induces only a sub-leading change on the gap at the Fermi surface at high 
   density\cite{Brown,Manuel}, but may have non-negligible effects at lower 
   densities. For treating  both the diagonal and off-diagonal parts of the 
   quark self-energy, the standard Eliashberg formalism must be used%
   \cite{Eliashberg,BCS}. If we include all the above effects, the gap equation 
   with frequency and momentum as independent variables should  be solved%
   \cite{2dgap}.

 Secondly, we assumed that there is no significant vacuum effects
    in the present paper. 
   If the density is close to the critical density of chiral symmetry 
     breaking, one must consider the interplay between
     the quark-antiquark condensate and the quark-quark condensate
     \cite{Berges-Rajagopal,Diakonov,Abuki}.
   This also requires us to treat the diagonal and off-diagonal components
     of the fermion self-energy simultaneously, which 
     corresponds to the Hartree-Fock-Bogoliubov theory in many-body 
     problem \cite{HFB}.

 Thirdly, we have   taken the Landau gauge $\xi=0$ in this paper.  
    As we discussed in Sect.~IID and in  Appendix A, the $\xi$-dependence
    in the gap equation remains  as far as we properly treat the 
    momentum dependence in the gap equation. Although it is a sub-leading 
    contribution relative to the magnetic and magnetic interactions, 
    it is desirable to develop a gauge invariant formalism especially 
    when one treats the low density region.

 Lastly, it will be very intriguing to find a new way of describing
     color superconductivity at low densities.
   The small size Cooper pairs with coherence length comparable
     to the inter-quark distance suggests that BEC description may
     be useful as in the analogous example in condensed matter physics.
   An analysis  along this line has been discussed in Ref.~\cite{RSSV_98}
    using the linear  sigma model for the diquark field.

\section*{Acknowledgments}

H.A. would like to thank T. Tatsumi for his useful
  comments at the early stages of the work.
He is also thankful to K. Suzuki for his various supports.
K.I. is grateful to R. Pisarski, D. Rischke, T. Sch\"afer, and D. T. Son
  for their critical comments and suggestions.
H.A. and K.I. are thankful to M. Matsuzaki for discussion.
Lastly, H.A. and T.H. acknowledge V.A. Miransky for his interests
  in the work.

\appendix
\section{Gauge dependence of $\Delta_+$}
In this Appendix, we discuss  the gauge parameter
 dependence of the gap equation for $\Delta_+$
  at high density.
 We write the gauge dependent contribution to 
 the gap equation for $\Delta_+(k)$
    as $I_{\xi} (k)$. Neglecting the antiquark pole, it becomes
\beq
  I_{\xi} ( k)&=&\frac{g^2}{24\pi^2}%
  \int_0^\infty dq\frac{q}{k}\int_{-1}^{1}d(\cos\theta)
    \frac{\Delta_+(q)}{\sqrt{E_+(q)^2+\Delta_+(q)^2}}%
     \  K_\xi(q,k;\cos\theta)  ,
\eeq
where $\hat{\q}\cdot\hat{\k}=\cos\theta$ and  $K_\xi$ is defined as
 \beq
   K_\xi(q,k;x)&=&\xi\times\frac{2qk}{q^2+k^2-2qk\cos\theta}%
  \left\{\frac{(q-k\cos\theta)(k-q\cos\theta)}{q^2+k^2-2qk\cos\theta}%
    -\frac{1-\cos\theta}{2}\right\}.
\eeq
$I_{\xi} ( k)$ should be added
    to the right-hand side  of Eq.~(\ref{full-gap}) when $\xi \neq 0$.
 
 Now,
 if we set   $q=k=\mu$ before the angular integration, the kernel 
  vanishes for all $\cos\theta$, namely  $K_\xi (\mu,\mu;\cos\theta)=0$.
 This is the standard argument that the gauge parameter dependence 
  does not appear in the gap equation at extremely high density.
 However, if we first integrate over $\cos\theta$, and then take the limit
  $q,k\to\mu$, the result is nonzero. This can be seen explicitly
   by carrying out  the angular integral of $K_\xi$ and 
    by writing the result in terms of  a variable
  $Y=(k/q+q/k)/2\ge 1$  :
\beq
  \int_{-1}^{1}d(\cos\theta) \ K_\xi(q,k;\cos\theta)%
    =\xi\left(\frac{Y-1}{2}\ln\left|\frac{Y+1}{Y-1}\right|-1\right).
\eeq
If we take the limit $Y\to 1$ in the right-hand side, the first term
  disappears, but the second term survives and gives a gauge-dependent
  contribution $-\xi$. One can trace back the origin of this
   situation by introducing a small regulator   $\e$ to 
    the collinear region of the integral  $\cos\theta\sim 1$.
  Then, the integral above becomes
\beq
  F(\e,Y)\equiv\int_{-1}^{1-\e}d(\cos\theta) K_\xi(q,k;\cos\theta)%
    =\xi\left(\frac{Y-1}{2}\ln\left|\frac{Y+1}{Y-1+\e}\right|
-\frac{Y-1}{Y-1+\e}\frac{Y+1}{2}\right).
\eeq
The first term  disappears irrespective of the order of $Y\to 1$ and
  $\e\to +0$.
However, the second  term vanishes only when one takes $Y\to 1$
  before $\e\to 0$.
Namely,
\beq
  \lim_{\e\to 0}\lim_{Y\to 1}F(\e,Y)&=&0,\\
  \lim_{Y\to 1}\lim_{\e\to 0}F(\e,Y)&=&\xi.
\eeq
This non-commutability of the two limits
 arises from the fact that  $F(\e,Y)$ does not approach
  to $  F(0,Y)$ uniformly  in the  region $Y\ge 1$.

Therefore, if we integrate over the angular variable $\cos\theta$ exactly,
     the  gap equation for $\Delta_+(k)$ at high density
   has an extra gauge-dependent contribution from the quark-pole;
   \beq
  I_\xi (k)&=&-\xi\frac{g^2}{24\pi^2}
    \int_0^\infty dq\frac{\Delta_+(q)}{\sqrt{E_+(q)^2+\Delta_+(q)^2}}
    \frac{q}{k}\left(1-\frac{(q-k)^2}{4qk}\ln\left|%
     \frac{(k+q)^2}{(k-q)^2}\right|\right).
\eeq
 Compared with the leading magnetic contribution with the kernel
  of logarithmic singularity, the above contribution is
   considered to be a sub-leading effect.

\section{Pippard length in the weak-coupling limit}
 In this Appendix, we derive the Pippard length  from the correlation function
     in the weak-coupling region.
 We use the two-point correlation function of quarks defined in
  Eq.~(\ref{correlation})
 \begin{eqnarray}
    \hat{\varphi}_+(q)&=& %
   \frac{\Delta_+(q)}{\sqrt{E_+(q)^2+\Delta_+(q)^2}},\label{eq:1}
 \end{eqnarray}
  Since   $\hat{\varphi}_+(q)$ is a function of $q (>0)$ only, its Fourier
   transformation is given by
 \beq
   \varphi_+(r)&=&\frac{N}{2\pi^2}\int_0^\infty %
  dq\,q^2 j_0(qr)\hat{\varphi}_+(q) \nonumber \\
   &=&\frac{N}{2\pi^2 r}%
        \int_0^\infty dq\,\sin(qr)\frac{q\Delta_+(q)}%
        {\sqrt{(q-\mu)^2+\Delta_+(q)^2}},
   \eeq
 where $j_0(x)=x^{-1}\sin x$ is the 0-th order spherical Bessel function.
 
 In the weak-coupling limit,  the integral
   is dominated at $q\sim\mu$. Then, by replacing $\Delta_+(q)$ by 
  $\Delta_+(\mu)$ and performing the $q$ integration approximately, 
  one finds
 \beq
   \varphi_+ (r)%
 &\sim&\frac{N}{2\pi^2 r}\int_0^\infty dq\,\sin(q r)%
        \frac{\mu\Delta_+(\mu )}{\sqrt{(q-\mu)^2+\Delta_+(\mu )^2}}\NN
       &\sim &\frac{N\mu^3}{\pi^2}\frac{\sin(\mu r)}%
 {\mu r}\frac{\Delta_+(\mu )}{\mu}
  K_0(\Delta_+(\mu) r),
 \eeq
 where $K_0$ is the 0-th modified Bessel function of the second kind.
 Using its asymptotic form 
 $$
  K_0(z)%
  \stackrel{z\to\infty}\sim
  \sqrt{\frac{\pi}{2z}}e^{-z}\sum_{n=0}^\infty%
  \frac{\Gamma(n+1/2)}{\Gamma(-n+1/2)}%
  \frac{1}{n!(2z)^n},
 $$
 we obtain
 \beq
 \varphi_+(r)%
  &\stackrel{r\to\infty}\sim&
   N\sqrt{\frac{\mu^5 \Delta_+(\mu)}{2 \pi^3}}%
   \frac{\sin \mu r}{(\mu r)^{3/2}}
               e^{-r/(\pi\xi_{\rm p})},
 \eeq
 where $\xi_{\rm p}\equiv (\pi\Delta_+(\mu))^{-1}$ is the Pippard length. 
 


\begin{thebibliography}{99}

\bibitem{Collins_Perry}
        J. C. Collins and M. J. Perry, Phys. Rev. Lett. {\bf 34}, 1353 (1975).
\bibitem{BL_84}
        D. Bailin and A. Love, Phys. Rept. {\bf 107}, 325 (1984).
\bibitem{Iwasaki}
	M. Iwasaki and T. Iwado, Phys. Lett. {\bf B 350}, 163 (1995);
        Prog. Theor. Phys. {\bf 94}, 1073 (1995).
\bibitem{ARW_98}
      	M. Alford, K. Rajagopal, and F. Wilczek, Phys. Lett. {\bf B422},
        247 (1998).
\bibitem{RSSV_98}
	R. Rapp, T. Sch\"afer, E. V. Shuryak, and M. Velkovsky,
	Phys. Rev. Lett. {\bf 81}, 53 (1998).
\bibitem{Review}
	For a recent review, see
	K. Rajagopal and F. Wilczek, ``{\it The Condensed Matter Physics
        of QCD}'', in ``At the frontier of particle physics -- 
	handbook of QCD'', Volume 3, Chapter 35, edited by M. Shifman 
        (World Scientific, 2001) [{\tt hep-ph/0011333}].
\bibitem{Schaefer}
	T. Sch\"afer and F. Wilczek, Phys. Rev. {\bf D60}, 114033 (1999).
\bibitem{Hong}
	D. K. Hong, V. A. Miransky, I. A. Shovkovy and L.C.R. Wijewardhana,
        Phys. Rev. {\bf D61}, 056001 (2000), Erratum-ibid. {\bf D62}
        (2000) 059903.
\bibitem{Pisarski-Rischke}
	R. D. Pisarski and D. H. Rischke, Phys. Rev. {\bf D61}, 074017
	(2000).
\bibitem{Son_98}
	D. T. Son, Phys. Rev. {\bf D 59}, 094019 (1999).
\bibitem{Berges-Rajagopal}
	J. Berges and K. Rajagopal, Nucl. Phys. {\bf B538}, 215 (1999).
\bibitem{Diakonov}
        G. W. Carter and D. Diakonov, Phys. Rev. {\bf D60} (1999) 016004.
\bibitem{BCS}
	J. R. Schrieffer, ``{\it Theory of Superconductivity}'',
	(Benjamin, New York, 1964)
\bibitem{Horie}
	R. Horie, ``{\it On Color Superconductivity in High Density
	Quark Matter}'', Master Thesis (Kyoto University, Feb. 1999),
\bibitem{Matsuzaki}
	M. Matsuzaki, Phys. Rev. {\bf D62}, 017501 (2000).
\bibitem{ARW_98b}
	M. Alford, K. Rajagopal, and F. Wilczek, Nucl. Phys. {\bf B537},
        443 (1999).
\bibitem{Higashijima}
        K. Higashijima, Phys. Rev. {\bf D29}, 1228 (1984);
        Prog. Theor. Phys. Suppl. {\bf 104}, 1 (1991).\\ 
         V. A. Miransky, Sov. J. Nucl. Phys. {\bf 38}, 280 (1983).
\bibitem{Aoki}
	K-I. Aoki, T. Kugo and M. Mitchard, Phys. Lett. {\bf B266}, 467 (1991).
	K-I. Aoki, M. Bando, T. Kugo and M. Mitchard,
	Prog. Theor. Phys.  {\bf 84}, 683 (1990).
\bibitem{Haymaker} 
        T. Kugo, ``{\it Basic Concepts in Dynamical Symmetry Breaking
        and Bound State Problems}'', lectures in 1991 Nagoya Spring School
        on Dynamical Symmetry Breaking (April 23-27, 1991, Nagoya, Japan)
        Ed. K. Yamawaki (World Scientific).\\
         V. A. Miransky, ``{\it Dynamical Symmetry Breaking in Quantum
      	Field Theories}'' (World Scientific, 1993).
\bibitem{LeBellac}
	M. LeBellac, ``{\it Thermal Field Theory}'' (Cambridge University
	Press, 1996).
\bibitem{Roberts}
    C. D. Roberts and S. M. Schmidt, 
     Prog.Part.Nucl.Phys. {\bf 45S1}, 1(2000).
\bibitem{Iida} K. Iida and G. Baym, Phys. Rev. {\bf D63}, 074018 (2001).
\bibitem{Pisarski-Rischke2}
    R. D. Pisarski and D. H. Rischke, Phys. Rev. {\bf D60}, 094013 (1999).
\bibitem{gauge_dependence}
	K. Rajagopal and E. Shuster, Phys. Rev. {\bf D62}, 085007 (2000).
\bibitem{Fetter-Walecka}
	A. L. Fetter and J. D. Walecka, ``{\it Quantum Theory of
        Many-Particle Systems}'' (McGraw-Hill, 1971).
\bibitem{BEC}
    	P. Nozi\'eres and S. Schmitt-Rink, J. Low Temp. Phys. {\bf 59},
        195 (1985). For recent references, see e.g., 
        E. Babaev, Int. J. Mod. Phys. {\bf A16}, 1175 (2001)
        and references thein. 
\bibitem{Brown}
	W. E. Brown, J. T. Liu, and H. C. Ren,
        Phys. Rev. {\bf D61}, 114012 (2000);
        {\it ibid.} {\bf D62}, 054013 (2000);
        {\it ibid.} {\bf D62}, 054016 (2000).
\bibitem{Manuel}
        C. Manuel, Phys. Rev. {\bf D62}, 114008 (2000). 
\bibitem{Eliashberg}
        See e.g.,
        D. J. Scalapino, in {\it Superconductivity}, edited by
        R. D. Parks (Dekker, New York, 1969).
\bibitem{2dgap}
        H. Abuki, T. Hatsuda, and K. Itakura, work in progress.
\bibitem{Abuki}
	H. Abuki, ``{\it Color Superconductivity in Quark Matter at High
        Density}'' Master Thesis (Kyoto University, Feb. 2000)
\bibitem{HFB} 
	P. Ring and P. Schuck, ``{\it Nuclear Many-Body Problem}''
 	(Springer, New York, 1980).
\end{thebibliography}
\end{document}